\documentclass{ws-mplb}

\usepackage{color}
\usepackage[normalem]{ulem}
\usepackage{graphicx}
\newcommand{\gbf}[1]{\boldsymbol #1}
\newcommand{\I}{{\rm i}}
\newcommand{\E}{{\rm e}}

\newcommand{\onlinecite}[1]{\protect\refcite{#1}}

\begin{document}

\markboth{Zi Yang Meng}{The characterization of topological properties in Quantum Monte Carlo simulations of the Kane-Mele-Hubbard model}

%
\catchline{}{}{}{}{}
%

\title{The characterization of topological properties in Quantum Monte Carlo simulations of the Kane-Mele-Hubbard model}

\author{Zi Yang Meng}

\address{
Department of Physics, University of Toronto, \\
Toronto, Ontario M5S 1A7,\\ Canada\\
Department of Physics \& Astronomy, Louisiana State University, \\
Baton Rouge, LA 70803,\\ USA\\
ziyangmeng@gmail.com}

\author{Hsiang-Hsuan Hung}

\address{Department of Physics, The University of Texas at Austin,\\
Austin, TX 78712, USA\\
hhhung@physics.utexas.edu}

\author{Thomas C. Lang}

\address{Department of Physics, Boston University,\\
Boston, MA 02215, USA\\
tcl@physics.bu.edu}

\maketitle

\begin{history}
\received{(Day Month Year)}
\revised{(Day Month Year)}
\end{history}

\begin{abstract}
Topological insulators present a bulk gap, but allow for
dissipationless spin transport along the edges. These exotic states
are characterized by the $Z_2$ topological invariant and are
protected by time-reversal symmetry. The Kane-Mele model is one model
to realize this topological class in two dimensions,
also called the quantum spin Hall state. In this review, we provide
a pedagogical introduction to the influence of correlation effects
in the quantum spin Hall states, with special focus on the
half-filled Kane-Mele-Hubbard model, solved by means of unbiased
determinant quantum Monte Carlo (QMC) simulations. We explain the
idea of identifying the topological insulator via $\pi$-flux
insertion, the $Z_2$ invariant and the associated behavior of the
zero-frequency Green's function, as well as the spin Chern number in
parameter-driven topological phase transitions. The examples
considered are two descendants of the Kane-Mele-Hubbard model, the
generalized and dimerized Kane-Mele-Hubbard model. From the $Z_2$
index, spin Chern numbers and the Green's function behavior, one can
observe that correlation effects induce shifts of the topological
phase boundaries. Although the implementation of these topological
quantities  has been successfully employed in QMC simulations to
describe the topological phase transition, we also point out their
limitations as well as suggest possible future directions in using numerical
methods to characterize topological properties of strongly
correlated condensed matter systems.
\end{abstract}

\keywords{Topological Insulator; Topological Invariants; Quantum Monte Carlo Simulation; Strongly Correlated Electrons.}

\tableofcontents

\section{Introduction}

The Ginzburg-Landau paradigm, the way to characterize
condensed matter states by means of spontaneously broken symmetries,
began to show its limitation in the past decades. The integer
quantum Hall (IQH) state constitutes a prominent example, where the
ground state of a two-dimensional electron gas, subjected to a
strong magnetic field, can no longer be characterized by symmetries
alone.\cite{Klitzing80} Although the quantum Hall state is an
insulator, it is topologically different from a trivial band
insulator because the ground states of these states cannot be
adiabatically connected to each other, unless the band gap
collapses. Moreover, there exist metallic states emerging on the
edges of the IQH sample.\cite{Halperin82} Such emergent chiral edge
modes also identify the distinction between the topological state
and a trivial band insulator. In the IQH, the Hall conductance
$\sigma_{xy}$ has been identified to be quantized, i.e., 
${\sigma_{xy} = n e^2/h}$ where $n$ is a nonzero
integer.\cite{Thouless82,Avron83} The integer number $n$ is the
topological invariant to identify the IQH state, also called the
Chern number or the TKNN number (stands for
Thouless-Kohmoto-Nightingale-Nijs).\cite{Thouless82} For
a trivial insulator, $n=0$. It defines the quantized conductance
with respect to the strength of the applied magnetic field, while
the symmetry of the ground state remains unchanged.

The IQH state is a member of the general class of symmetry protected
topological (SPT) phases with a short-range entangled ground
state,\cite{Wen2013} which edges states are protected by charge-
and spin-$S_z$ invariance, while time reversal symmetry (TRS) is
broken due to the external magnetic field. Topological insulators,
constitute another subgroup that cannot be
classified within the Ginzburg-Landau
paradigm.\cite{Kane05a,Kane05b,Xu2006,Qi2006,Bernevig2006,Koenig2007,Fu2007prb,Fu2007b,Zhang2009,Chen2009}
Different from the IQH, these states preserve their particle number
and TRS, and can be realized experimentally without the need of a
magnetic field.\cite{Koenig2007,Chen2009,Roy2013} In these systems,
spin-orbital interactions play a key role as an effective magnetic
field for each spin species. The quantum spin Hall state (QSH) is a
two-dimensional version of a topological insulator and was
theoretically proposed in the context of graphene, called the
Kane-Mele (KM) model\cite{Kane05a,Kane05b} and in the HgTe quantum
wells described by the Bernevig-Hughes-Zhang
model.\cite{Bernevig2006} In this review, we focus our discussion on
the former. For the latter case, we refer the reader to the
Refs.~\onlinecite{konig2008review,maciejko2011}. In their
seminal papers\cite{Kane05a,Kane05b}, Kane and Mele
show that the intrinsic spin-orbit coupling opens a bulk gap,
and leads to the emergence of robust helical edge states.
These helical states consist of two spin channels,
each of which carries opposite chirality and is protected by TRS
against non-magnetic impurities.\cite{Wu06} In contrast to the IQH
state where the TKNN number ${n \in \mathbb{Z}}$ can be any integer, the
topological index of the QSH state, denoted as $\nu$, is in the
$Z_2$ symmetry class, i.e., $\nu=0,1$.

Generalizing the non-interacting KM model to the more realistic case
of interacting electrons raises the following questions: how do
electronic correlations affect the topological phase? Does the
topological state remain stable under correlations? Investigations to
answer this question in the KM model with correlations have been performed by means of
the mean-field theory,\cite{Rachel2010} Schwinger Boson
approach,\cite{vaezi2012} variational Monte Carlo,\cite{yamaji2011}
cellular dynamical mean field theory,\cite{wuwei2012} variational
cluster approximation\cite{shunliyu2011} and determinant quantum
Monte Carlo (QMC) simulations.\cite{Lang2011,Zheng2011,Hohenadler2012} In this 
brief review, we are trying to provide a pedagogical introduction to
classify the complex interplay between the topological insulator and
electron correlations by means of $\pi$-flux insertion, the $Z_2$
topological invariant and the spin Chern number. Our focus lies specifically on the
implementation using unbiased and numerically exact auxiliary field QMC
simulations of the interaction version of the KM model, the
Kane-Mele-Hubbard (KMH) model, and the resulting physical
consequences, such as the correlated QSH state and the relation to
the $Z_2$ topological invariant. For more general reviews and
articles on topological insulators, we encourage readers to look
into the Refs.~\onlinecite{konig2008review,maciejko2011,moore08,Hasan2010,Qi2010phystoday,Qi2011,hasan2011,Fiete2012,Hohenadler2013,xidai2012,ando2013}.

In the following, we first explain the generic ingredients of
characterizing the topological quantum phase transitions by means of
magnetic flux insertion in the Kane-Mele-Hubbard (KMH)
model,\cite{Assaad2013} which very effectively allows to test for
emerging edge states. We then introduce the $Z_2$ invariant which is
used to characterize the change of the time-reversal polarization
due to a flux quantum $h/2e$ threading through a
torus.\cite{Fu2006prb} It follows the description of the evaluation
of the $Z_2$ topological invariants in terms of
eigenstates of tight-binding Hamiltonian in the noninteracting
limit. With the inversion symmetry, the $Z_2$ evaluation can be
associated with the parity of the eigenstates at the time-reversal
invariant momenta (TRIM).\cite{Fu2007prb}. The
formalism of the $Z_2$ index is then extended to the interacting case, 
where the zero-frequency Green's function plays an essential role for 
the topological invariants.\cite{Wang2012prx} We introduce the QMC
algorithm which allows us to accurately acquire the interacting Green's 
function, provide examples in two descendants of the KMH model:
The generalized and dimerized KMH models, for which we study the
topological properties under the influence of the local Hubbard
interaction.\cite{Hung2013b,Hung2013,Lang2013} 
In our numerical results, we discovered that the correlation can
stabilize, or destabilize the topological insulators, and the
parameter-driven topological phase transitions can be described by
the $Z_2$ topological invariant at the interacting level. We also discuss
possible limitations of the $Z_2$ topological invariant for 
interaction-driven phase transitions using QMC simulations.
Furthermore, we introduce the concept of the spin Chern number and its
effective implementation\cite{Hung2013b} as another successful approach
to determine the topological nature of phases in QMC simulations.

\section{The Kane-Mele-Hubbard Model}
\label{sect:kanemelemodel}

\subsection{The Kane-Mele Model: a Quantum Spin Hall Insulator}
\label{sect:QSHI}

\begin{figure}[tp]
\centering
\includegraphics[width=\textwidth]{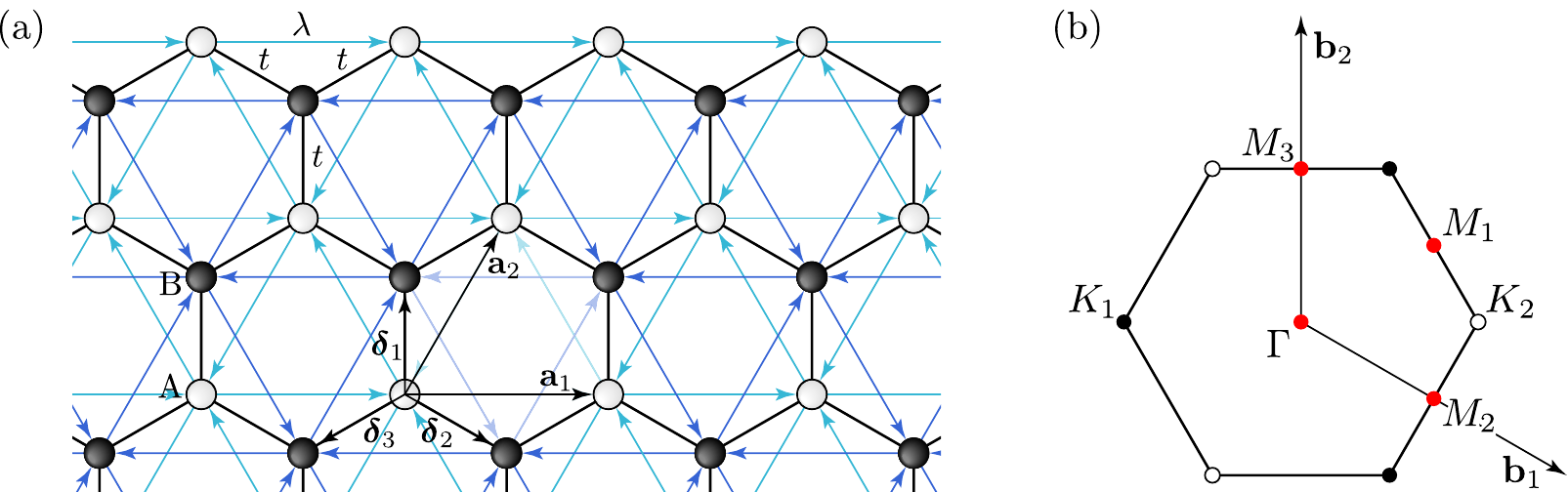}
\caption{(a)
The Kane-Mele-Hubbard model for one spin species with a zig-zag
edge. The underlying honeycomb lattice consists of two sublattices
A, B, denoted by the filled and open circles, and is spanned by the
primitive vectors $\mathbf{a}_1=(\sqrt{3},0)$,
$\mathbf{a}_2=(\frac{\sqrt{3}}{2},\frac{3}{2})$; the bond length is set to
be unity. The nearest-neighbor hopping $t$ connects lattice sites
belong to different sublattices by the vectors
$\gbf{\delta}_1=(0,1)$, $\gbf{\delta}_{2,3}=(\pm
\frac{\sqrt{3}}{2},-\frac{1}{2})$. Complex values
next-nearest-neighbor spin-dependent hopping $\I\lambda$ connects
lattices sites within the same sublattice. The left (right) turn
next-nearest-neighbor hopping is associated with $v_{ij}=1$ ($-1$)
in the Hamiltonian. (b) The Brillouin zone of the honeycomb lattice.
Filled and open circles denote the Dirac points $K_{1,2}=(\pm
\frac{4\pi}{3\sqrt{3}},0)$ and the time-reversal invariant momenta
(TRIM) denoted by the red circles are $\Gamma=(0,0)$, $M_{1,2}=(
\frac{\pi}{\sqrt{3}},\pm\frac{\pi}{3})$, and $M_3=(0,\frac{2\pi}{3})$.
$\gbf{b}_{1}=(\frac{2\pi}{\sqrt{3}},-\frac{2\pi}{3})$
and $\gbf{b}_{2}=(0,\frac{4\pi}{3})$ are reciprocal vectors.
\label{fig:KMmodel} }
\end{figure}

The Kane-Mele (KM) model was derived as a model with an intrinsic 
spin-orbital interaction on a two-dimensional honeycomb lattice -- 
the structure of graphene. The model with its lattice structure and 
parameters is illustrated in Fig.~\ref{fig:KMmodel}(a). The idea proposed 
by C. Kane and E. Mele was to construct a spinful model which consists 
of two copies of the the Haldane model\cite{Haldane1988} with opposite 
spin.\cite{Kane05a,Kane05b} Although the spinless Haldane model alone 
breaks TRS, the spinful KM model is time-reversal invariant. The 
Hamiltonian of the KM model is given by,
\begin{equation}
   H_\text{KM} = -t\sum_{\langle i,j\rangle,\sigma}c^{\dagger}_{i\sigma}c_{j\sigma} +
                 \I\,\lambda\sum_{\langle\!\langle i,j \rangle\!\rangle, \alpha\beta} v_{ij}\,c^{\dagger}_{i\alpha}\sigma^{z}_{\alpha\beta}c_{j\beta}\;,
\label{eq:KaneMeleHamiltonian}
\end{equation}
where $\alpha$, $\beta$, $\sigma$ denote the spin species $\uparrow$ and
$\downarrow$. The first term describes the nearest
neighbor hopping on a honeycomb lattice. The second term represents
spin-orbit coupling, it connects next-nearest-neighbor sites with a
complex (time-reversal symmetric) hopping with amplitude $\lambda$.
The factor $v_{ij}=-v_{ji}=\pm1$ depends on the orientation of the
three nearest neighbor bonds the electron traverses in going from
site $j$ to $i$ and affects the orientation of the
next-nearest-neighbor bonds for one spin species. As shown in
Fig.~\ref{fig:KMmodel}(a), $\nu_{i,j}=\pm1$ if the electron makes a
left (right) turn to get to the next-nearest-neighbor site. The
$\sigma^{z}_{\alpha\beta}$ in the spin-orbit coupling term is the
z-component of the Pauli matrix, which furthermore distinguishes the
$\uparrow$ and $\downarrow$ spin states with opposite
next-nearest-neighbor hopping amplitude; thus the
next-nearest-neighbor hopping is spin-dependent. Physically, the
$\lambda$ term stands for the intrinsic spin-orbit coupling, where
$S^{z}$ is conserved, and amounts to a staggered spin-dependent
magnetic field threading the triangular plaquettes defined by the
next-nearest-neighbor bonds.

To better understand the KM model, we Fourier transform the
Hamiltonian in Eq.~(\ref{eq:KaneMeleHamiltonian}) into momentum
space. In terms of the spinor
${\Phi^{\dagger}_{\mathbf{k}}=(c^{\dagger}_{{\rm A},\mathbf{k},\uparrow},
c^{\dagger}_{{\rm B},\mathbf{k},\uparrow},
c^{\dagger}_{{\rm A},\mathbf{k},\downarrow},
c^{\dagger}_{{\rm B},\mathbf{k},\downarrow})}$,
Eq.~(\ref{eq:KaneMeleHamiltonian}) is recast as
$H_\text{KM}=\sum_{\mathbf{k}}\Phi^{\dagger}_{\mathbf{k}}H(\mathbf{k})\Phi_{\mathbf{k}}$
in basis of $(\uparrow,\downarrow)\otimes({\rm A},{\rm B})$ and expressed in a
block-diagonal form as
\begin{equation}
 H(\mathbf{k}) = \left( \begin{array}{cccc}
\gamma_{\mathbf{k}} & -g_{\mathbf{k}} & 0 & 0 \\
-g^{\ast}_{\mathbf{k}} & -\gamma_{\mathbf{k}} & 0 & 0 \\
0 & 0 & -\gamma_{\mathbf{k}} & -g_{\mathbf{k}} \\
0 & 0 & -g^{\ast}_{\mathbf{k}} & \gamma_{\mathbf{k}} \end{array}
\right) = \left( \begin{array}{cc}
H_{\uparrow}(\mathbf{k}) & 0 \\
0 & H_{\downarrow}(\mathbf{k}) \end{array} \right)\;.
\label{eq:kmhamiltonianmatrix}
\end{equation}
Here, ${g_{\mathbf{k}}=t\sum^{3}_{i=1}
\E^{\I\mathbf{k}\cdot\gbf{\delta}_{i}}}$ comes from the nearest
neighbor hopping and
${\gamma_{\mathbf{k}}=2\lambda[2\cos(3k_{y}/2)\sin(\sqrt{3}k_x/2)-\sin(\sqrt{3}k_x)]}$
represents the spin-orbit interaction. Each of the block diagonal
matrices $H_{\sigma}$ represents a Haldane model for one spin
species.\cite{Haldane1988} Although $H_{\sigma}$ individually breaks
the TRS, the whole Hamiltonian in Eq.~(\ref{eq:kmhamiltonianmatrix})
recovers it at the time-reversal invariant momenta $\{\Gamma, M_{0},
M_{1}, M_{2}\}$. The argument is given as follows:
consider the time-reversal operator
$\mathcal{T}=\E^{\I\pi\gbf{\sigma}^y}\mathcal{K}$, where
$\gbf{\sigma}^y$ is the Pauli matrix applied in the sublattice space
and $\mathcal{K}$ denotes the complex conjugation
operator.\cite{Fu2007prb,Rachel2010} Application of $\mathcal{T}$ to
a single-particle Bloch state means to invert the momentum
from $\mathbf{k}$ to $-\mathbf{k}$, flip the spin from $\uparrow$ to
$\downarrow$, and the complex conjugate is to be taken. By
interchanging the $\uparrow$ and $\downarrow$ sectors of
Eq.~(\ref{eq:kmhamiltonianmatrix}), taking the complex conjugate and
considering ${\gamma_{-\mathbf{k}}=-\gamma_{\mathbf{k}}}$, as well
as ${g_{-\mathbf{k}}=g^{\ast}_{\mathbf{k}}}$, one can  verify that
${H(-\mathbf{k})=\mathcal{T}H(\mathbf{k})\mathcal{T}^{-1}}$. Thus,
the KM model is time-reversal invariant only while
$\mathbf{k}=-\mathbf{k}$, i.e., at the TRIM.

\begin{figure}[tp]
\centering
\includegraphics[width=0.9\textwidth]{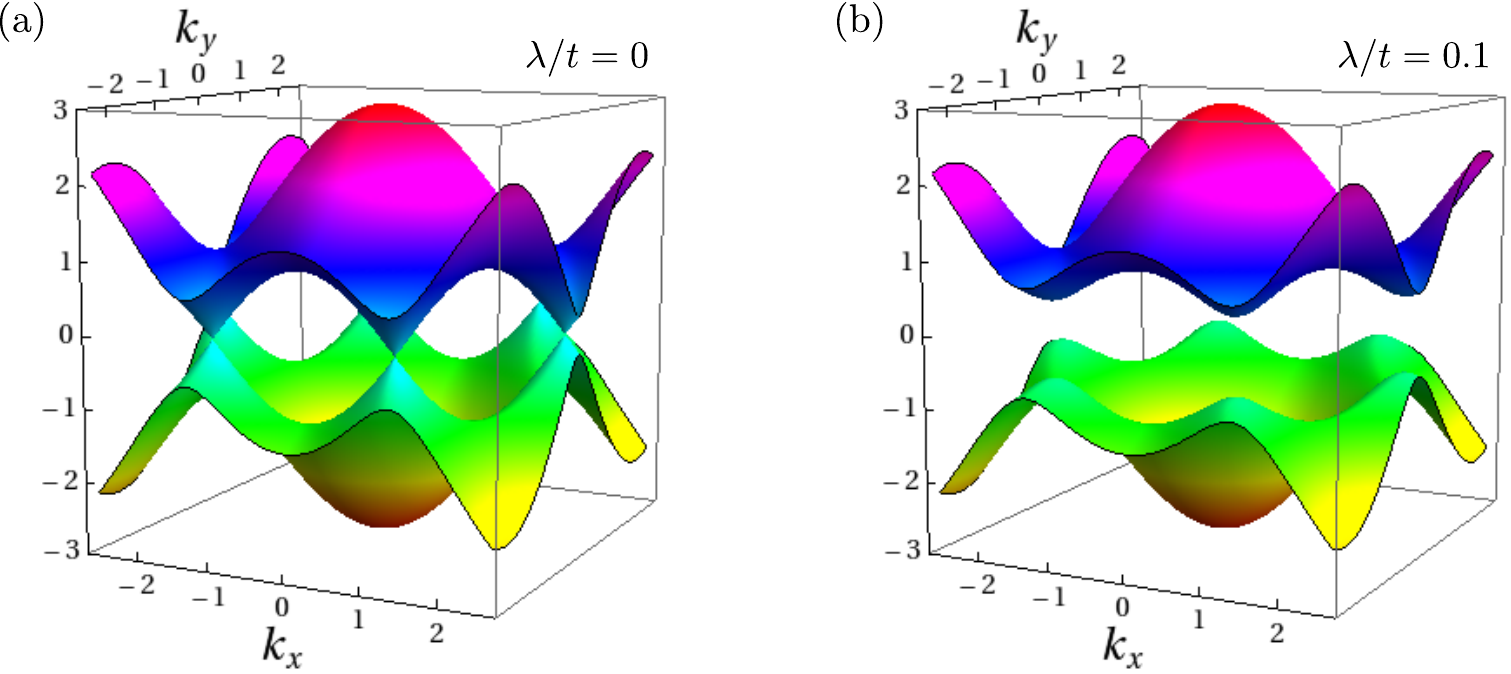}
\caption{The
free dispersions of the KM (a) at spin-orbit coupling $\lambda=0$, where there are six
(two distinct) Dirac points at $K_{1,2}=(\pm
\frac{4\pi}{3\sqrt{3}},0)$ and the system is a semi-metal (b) at
$\lambda/t=0.1$, where the spin-orbit coupling opens a gap $\Delta_\text{so}/t=3\sqrt{3}$ at the Dirac
points and the system is a quantum spin Hall insulator.
\label{fig:kmtopologicalphasetransition}}
\end{figure}

At $\lambda=0$, Eq.~(\ref{eq:kmhamiltonianmatrix}) gives rise to the
famous graphene band dispersion
${\varepsilon(\mathbf{k})=\pm|g_{\mathbf{k}}|=\pm
t[3+2\cos(\sqrt{3}k_x)+4\cos(3k_y/2)\cos(\sqrt{3}k_x/2)]^{1/2}}$, in
which the conduction bands and valence bands touch at the Dirac
points $K_{1,2}$
[filled and open circales in Fig.~\ref{fig:KMmodel}(b)]. Around the
Dirac points the band dispersion is linear and forms Dirac cones. The band structure of
the graphene dispersion is depicted in
Fig.~\ref{fig:kmtopologicalphasetransition}(a). At half-filling,
i.e., the number of electrons equals the number of lattice sites,
the Fermi level is located exactly at the Dirac points and the
system is gapless with a vanishing density of states and is hence a
semi-metal.\cite{CastroNeto2009}

Any finite $\lambda$ opens a bulk gap.
Figure~\ref{fig:kmtopologicalphasetransition}(b) show the case of
$\lambda=0.1t$. Since the KM model Hamiltonian can be decoupled as
two independent Hamiltonian $H^{\sigma}(\mathbf{k})$, the dispersion
of the KM model can be easily solved as
${\varepsilon_\text{KM}(\mathbf{k})=\pm\sqrt{|g_{\mathbf{k}}|^{2}+\gamma_{\mathbf{k}}^{2}}}$,
each of them is double degenerate. The bulk gap at the Dirac points
opens as ${\Delta_\text{so}=3\sqrt{3}\lambda t}$.\cite{Rachel2010}
Note that the inversion symmetry breaking field, e.g.,
a staggered potential term ${\sum_{i,\sigma} \varepsilon_i
c^{\dag}_{\sigma}c_{\sigma}}$, where ${\varepsilon_i = 1}$ ($-1$) on
sublattice A (B), also opens a gap. Hence
topological trivial and nontrivial insulators cannot be easily
distinguished by the bulk gaps. The KM model however features the
hallmark of protected edge states once a boundary is introduced into
the system according to the bulk-edge-correspondence in non-trivial
topological systems.\cite{Hasan2010,Qi2011} The edge state of the KM
model can be seen by solving Eq.~(\ref{eq:KaneMeleHamiltonian}) on a
ribbon geometry.

\begin{figure}[tp]
\centering
\includegraphics[width=8cm]{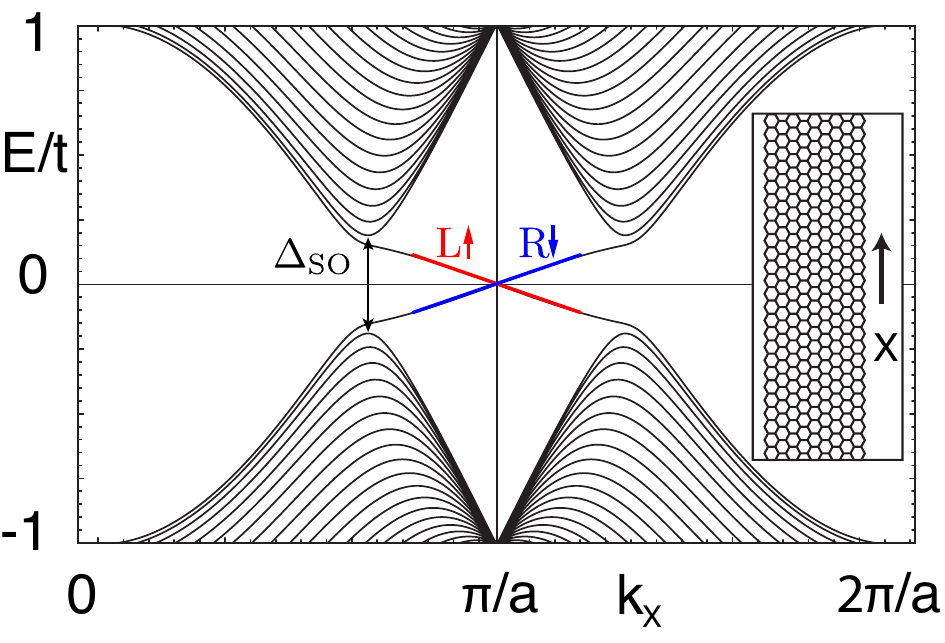}
\caption{One
dimensional energy bands for a ribbon geometry (zigzag ribbon, see
inset) of the KM model at ${\lambda = 0.03t}$. The edge states
traverses the bulk gap at the $K_1$ and $K_2$ points. Due to the
spin-orbit coupling, the momentum of the edge states is on lock with
their spin states. For example, the leftmover $L$ is in the
$\uparrow$ spin state and the right mover $R$ is in the $\downarrow$
spin state. The number of pairs of edge states corresponds to the
$Z_2$ topological invariant. Adapted and reproduced with permission
from Ref.~\onlinecite{Kane05b}. Copyright 2005 American Physical
Society. \label{f2}}
\end{figure}

Figure~\ref{f2} shows the one-dimensional band structure for a
zigzag ribbon (as shown in the inset).\cite{Kane05a,Kane2007} In the
projection onto the one-dimensional edge, one can see the bulk band
gap $\Delta_\text{so}$ at the $K_1$ and $K_2$ points as indicated by
the arrows. There exist two bands within the band gap, which connect
the $K_1$ and $K_2$ points. These transverse modes are states
localized on the edges of the zigzag ribbon, which is analogous to
the chiral modes localized on the edge of the IQH state. Here
however, the electrons with $L\uparrow$ and $R\downarrow$ spin
states propagate in opposite directions along one edge; thus it is
bidirectional and the net charge carrier is zero. The bidirectional
channels however bring a nonzero spin current ${J_{s} =
\frac{\hbar}{2e}(J_{\uparrow}-J_{\downarrow})}$,\cite{Kane05b} and
the spin Hall conductivity characterizes that the magnitude of spin
currents are carried by the opposite spin components on the edges.
The edge states are named helical state\cite{Wu06} and are
essentially one dimensional chiral Dirac fermions which occurrence
is contingent on the properties of the two dimensional bulk system.
This so-called bulk-boundary correspondence states the fact that the
existence of edge states is guaranteed by the topological nature of
the bulk system and the two are inextricably linked with each other. 
This spin-filtered state is topologically different
from an ordinary one-dimensional metal, where the electronic
behavior is not spin-filtered.\cite{Kane2007} Note that the helical
states cross at $k_x=\pi$, and are hence protected by the
time-reversal symmetry. This means that the edge states are also robust
against time-reversal symmetric impurities.\cite{Wu06} The number of
pairs of edge states (modulo 2) is directly linked to the value of
the $Z_2$ topological invariant $\nu$.\cite{Kane05a,Hasan2010} The
protection of the topological state with respect to adiabatic
deformations implies that the only way to achieve a change of the
topological invariant is to close the bulk band gap. Hence,
investigation of edge properties can be used to identify the
corresponding properties of the bulk. Note that the statement above
is valid as long as the invariant is well defined. Indeed, the
topological properties of a system can be changed without closing
the bulk gap in the single-particle
spectrum.\cite{Ezawa2013,yang2013,rachel2013} However, in
an interacting system beyond the mean-field approach
the spontaneous symmetry breaking associated with a direct
transition from a topological insulator to a topologically trivial
phase is always accompanied by the closing of a gap, albeit in the
charge-, or spin-sectors.\cite{Hohenadler2012}

As mentioned previously, Eq.~(\ref{eq:kmhamiltonianmatrix}) consists
of two Haldane model copies for each spin species, and each of them
provides an IQH state with the quantized Hall conductivity
$\sigma_{xy}=\pm e^2/h$. In close analogy to the IQH effect, the QSH
insulator has a quantized spin Hall conductivity,
$\sigma^s_{xy}=e/2\pi$, showing the nontrivial topological feature.
Each spin species contributes a nontrivial Chern number
$C_{\sigma}$. The time-reversal symmetry guarantees that the Chern
numbers for the two spin sectors have opposite signs
$C_{\uparrow}=-C_{\downarrow}=1$. Therefore, we have the net charge
Chern number $C_c=C_{\uparrow}+C_{\downarrow}=0$, but a nonzero
$C_s=(C_{\uparrow}-C_{\downarrow})/2 \ne 0$, which defines the quantized
spin Hall conductivity $\sigma^s_{xy}$ in terms of $e/2\pi$ and has
been shown to be robust against time-reversal symmetric disorder and
magnetic
Kondo-Impurities.\cite{Kane2007,Sheng2005,Sheng2006,Goth2013}
Although a nonzero value in $C_s$ indicates the nontrivial
topological property, $\sigma^s_{xy}$ is however not
necessarily quantized.\cite{Kane05b,Kane2007} For example, in the presence of the Rashba
coupling
\begin{equation}
H_\text{R} = \I\,\lambda_\text{R} \sum_{\langle i,j\rangle,\alpha
\beta}c^{\dag}_{i,\alpha}\Big[ \hat{z}\cdot (\gbf{\sigma} \times
\mathbf{d}_{ij}) \Big]_{\alpha \beta}c_{j,\beta}\;, \quad\mbox{with}\quad \mathbf{d}_{ij}=\gbf{\delta}_{1,2,3}\;,
\end{equation}
which preserves the time-reversal symmetry, but breaks inversion
symmetry and causes $S^z$ to be no longer conserved, the spin Hall
conductivity will deviate from quantization and can take continuous
values.\cite{Sheng2005,Sheng2006} Nevertheless, as long as the band
gap remains open, the system with finite $\lambda_\text{R}$ remains
a QSH insulator. Consequently, the spin Hall conductivity does not
constitute a topological invariant for the QSH state. Instead, the
topological invariant of the QSH state is the $Z_2$ invariant, which
is given by
\begin{equation}
   \nu =C_{s} \ \text{mod} \ 2\;,
\end{equation}
and takes on the values of 0 or 1. The value $\nu=1$ corresponds to
the topologically nontrivial QSH state and $\nu=0$ corresponds to a
topologically trivial insulator.\cite{Kane05b,Hasan2010,Qi2011} The
authors of Refs.~\onlinecite{Kane05b,Kane2007} have shown that even in
the presence of finite $\lambda_\text{R}$ (below the threshold value
which  closes the bulk gap), $\nu=1$ and therefore the $Z_2$
invariant is indeed a proper description to distinguish the QSH
regime from a topologically trivial insulator.

\subsection{Quantum Monte Carlo Simulations \& the Kane-Mele-Hubbard Model}
\label{sect:QMCforKMH}

Next let us turn to the QSH state under the influence of
interactions. The KM model in Eq.~(\ref{eq:KaneMeleHamiltonian}) is
non-interacting. To consider electron interactions, the simplest
non-trivial approach is to augment the KM model by an additional
on-site Coulomb repulsion of strength $U$, which results in the
Kane-Mele-Hubbard (KMH) model given by
\begin{equation}
H_\text{KMH} = -t\sum_{\langle i,j
\rangle,\sigma}c^{\dagger}_{i\sigma}c_{j\sigma} +
\I\,\lambda\sum_{\langle\!\langle i,j \rangle\!\rangle, \alpha\beta}
v_{ij}\,c^{\dagger}_{i\alpha}\sigma^{z}_{\alpha\beta}c_{j\beta} +
\frac{U}{2}\sum_{i}(n_i-1)^2\;. \label{eq:KaneMeleHubbardHamiltonian}
\end{equation}
The KMH model is a many-body Hamiltonian which can no longer be
diagonalized via a Fourier transformation. In order to study the
topological nature of the KMH model in the presence of
interaction, we need to employ more advanced techniques. At
half-filling, the bipartite nature of the KMH model
sports particle-hole symmetry and TRS, so that QMC
simulations can be employed to solve this system in a controlled and
unbiased way on large lattices. In this section, we briefly introduce the determinant
QMC technique which is used to study the KMH models in following
sections. For more detailed description, we encourage readers to
refer more specific articles.\cite{Assaad2002,Assaad2008}

The determinant QMC has been shown to be an excellent and unbiased approach 
to deal with strongly correlated system with Hubbard interactions.\cite{Sugiyama1986,Sorella1989,White1989,Meng2010,Sorella2012} 
In the zero temperature (${T=0})$ projector algorithm, the ground state wave function 
$|\Psi_0\rangle$ can be obtained by stochastic projection of the Hamiltonian onto a trivial wave function 
$|\Psi_\text{T}\rangle$, provided a finite overlap $\langle \Psi_\text{T} 
|\Psi_0\rangle \ne 0$. For the KMH model, the lowest 
single-particle state of $H_\text{KM}$ is a good candidate for the trial 
wave function $|\Psi_\text{T}\rangle$. The expectation value of an arbitrary 
observables $O$ is obtained by
\begin{eqnarray}
   \langle O \rangle =\lim_{\Theta \to \infty}\frac{\langle
\Psi_\text{T} |\E^{- \frac{\Theta }{2}H} O \E^{- \frac{\Theta
}{2}H}|\Psi_\text{T}\rangle}{\langle \Psi_\text{T} | \E^{-\Theta H}| \Psi_\text{T}\rangle}\;.
\label{eqn:expection}
\end{eqnarray}
The imaginary time axes is discretized into $M$ Trotter-slices such that 
the projection operator $\E^{-\Theta H}=[\E^{-\Delta \tau H}]^M$ for 
$M\to\infty$, with the projection length $\Theta=\Delta\tau M$ and 
$\Delta\tau \ll 1$. Using the first order Suzuki-Trotter decomposition, 
$\E^{-\Delta \tau H}$ can be decomposed as
\begin{eqnarray}
  \E^{-\Delta \tau H} \simeq \E^{-\Delta \tau H_\text{KM}} \E^{-\Delta \tau H_U} \;.
\label{eqn:suzuki}
\end{eqnarray}
The interaction term $H_U$ is non-bilinear in the fermionic operators 
and cannot be expressed in the single-particle basis. However, the 
discrete SU($N$)-invariant Hubbard-Stratonovich transformation,\cite{Assaad2008} 
allows to transform the interacting imaginary time-evolution operator 
$\E^{-\Delta \tau H_U}$ into bilinear form at the cost of the 
integration over a four-component auxiliary field $\ell$ on all sites.
\begin{eqnarray}
\E^{-\Delta \tau
\frac{U}{2}(n_i-1)^2}=\frac{1}{4}\sum_{\ell=\pm1,\pm2}\gamma(\ell_i)\,\E^{\sqrt{-\Delta
\tau U/2}\,\eta(\ell_i)(n_i-1)}+O(\Delta
\tau^4)\;,\label{eqn:HStransformation}
\end{eqnarray}
where
\begin{eqnarray}
   \gamma(\pm1) & = \;\; (1+\sqrt{6}/3)\;,\quad\quad\eta(\pm1) & = \;\; \pm\sqrt{2(3-\sqrt{6})}\;,\nonumber\\
   \gamma(\pm2) & = \;\; (1-\sqrt{6}/3)\;,\quad\quad\eta(\pm2) & = \;\; \pm\sqrt{2(3+\sqrt{6})}\;.
\end{eqnarray}
The systematic error of the Hubbard-Stratonovich transformation of
order $\Delta\tau^4$ is still small compared to $\Delta\tau^3$ error
introduced by the asymmetric Suzuki-Trotter decomposition Eq.
(\ref{eqn:suzuki}) and can be controlled by choosing appropriately
small values for $\Delta\tau$. In most of the QMC simulations in the
upcoming sections, we employ $\Delta\tau t=0.05$.

The integration over all auxiliary field configurations
$\ell_{i\tau}$ is performed using stochastic Monte Carlo sampling.
The partition function is given by
\begin{eqnarray}
   \langle\Psi_0|\Psi_0\rangle
   & = & \langle\Psi_{\text{T}}|\E^{-\Theta H}|\Psi_{\text{T}}\rangle = \langle\Psi_{\text{T}}|\prod_{\tau=1}^{M} \E^{-\Delta\tau H_\text{KM}} \E^{-\Delta\tau H_{U}}|\Psi_{\text{T}}\rangle \nonumber\\
   & = & \mbox{Tr}\left[
      \lim_{\Theta\to\infty} \E^{-\Theta(H_\text{T}-E_\text{T})} \prod_{\tau=1}^{M} \E^{-\Delta\tau H_\text{KM}} \E^{-\Delta\tau H_{U}}
   \right]\nonumber\\
   & = & \lim_{\Theta\to\infty}\sum_{\{\ell_{i\tau}\}}\prod_{i,\tau}\; \gamma(\ell_{i\tau})\,\prod_{\sigma}w_{\sigma}(\ell_{i\tau})\;.
   \label{Trace}
\end{eqnarray}
Here, ${|\Psi_\text{T}\rangle}$ a trial wave function which
corresponds to the non-degenerate ground state of a single-paritle
Hamiltonian $H_\text{T}$ with
${|\Psi_\text{T}\rangle\langle\Psi_\text{T}|=\lim_{\Theta\to\infty}
\E^{-\Theta(H_\text{T}-E_\text{T})}}$, where $E_\text{T}$ is the
corresponding non-degenerate ground state energy. We usually choose
$H_\text{T}=H_\text{KM}(\Phi)$, with $\Phi$ being a statistically irrelevant 
small magnetic flux threading the KM model on the torus in order to lift 
its degeneracy.\cite{Assaad2002,Assaad2008,Hohenadler2012} 
The sum $\sum_{\{\ell_{i\tau}\}}$ runs over possible auxiliary
configurations $\ell_{i\tau}$, where ${i = 1 \ldots N}$, ${\tau = 1 \ldots M}$.
The weight explicitly reads
\begin{equation}
   w_{\sigma} = \mbox{Tr}\left[\E^{-\Theta(H_\text{T}-E_\text{T})}\,
            \E^{-\Delta\tau\sum_{i,j} c^{\dagger}_{i\sigma} [H^{\sigma}_\text{KM}]_{ij} c_{j\sigma}}
            \E^{\alpha\sum_i\eta(\ell_{i\tau})(n_{i\sigma}-\frac{1}{2})}
      \right]\;,
      \label{weightsigma}
\end{equation}
with ${\alpha = \sqrt{-\Delta\tau U/2} = \I \alpha'}$ and $\alpha'=\sqrt{\Delta\tau U/2}$ for ${U>0}$.

In order to have QMC simulations free of the negative sign problem the configuration weights
$\prod_{\sigma}w_{\sigma}(\ell_{i\tau})$ must remain positive definite. In the half-filled KMH model,
TRS and particle-hole symmetry yield the condition $w_{\sigma} = w^{*}_{\bar{\sigma}}$.
To demonstrate this is fulfilled in the KMH model one just needs to check that the nearest-neighbor
hopping matrix elements, the spin-orbit hopping matrix elements and
the interaction matrix elements in Eq.~(\ref{Trace}) indeed render
such a property. As for the nearest-neighbor hopping, it is
spin independent, such that 
\begin{equation}
c_{i\sigma}^{\dagger} c_{j\sigma}^{\phantom{\dagger}} =
c_{i\bar{\sigma}}^{\dagger} c_{j\bar{\sigma}}^{\phantom{\dagger}}\;,
\end{equation}
and bipartite hopping matrix elements are real numbers and will automatically
give $w_{\sigma}=w^{*}_{\bar{\sigma}}$ if there are no other terms in the Hamiltonian.
The spin-orbit hopping matrix elements have a complex hopping
amplitude, but are complex conjugate with respect to $\sigma$ by construction, hence 
\begin{equation}
       \I\, c_{i\sigma}^{\dagger} c_{j\sigma}^{\phantom{\dagger}} = -\I\, c_{i\bar{\sigma}}^{\dagger} c_{j\bar{\sigma}}\;,
\end{equation}
satisfies the condition as well. At half filling $n_{\sigma}=1-n_{\bar{\sigma}}$, so that the interaction term has the same property:
\begin{equation}
\I\,\alpha'\,\eta(\ell_{i\tau})\Big[n_{i\sigma}-\frac{1}{2}\Big]=\I\,\alpha'\,\eta(\ell_{i\tau})\Big[1-n_{i\bar{\sigma}}-\frac{1}{2}\Big]=-\I\,\alpha'\,\eta(\ell_{i\tau})\Big[n_{i\bar{\sigma}}-\frac{1}{2}\Big]\;.
\end{equation}
Hence, the interaction matrix element for spin $\sigma$ is the
complex conjugate of the interaction matrix element for the other
spin $\bar{\sigma}$. Consequently, one can readily see that the
nearest neighbor hopping matrix elements, the spin-orbit
hopping matrix elements and the interaction matrix elements all
guarantee $w_{\sigma}=w^{*}_{\bar{\sigma}}$ and hence the
configurational weight
$\prod_{\sigma}w_{\sigma}(\ell_{i\tau})=w_{\sigma}(\ell_{i\tau})w_{\bar{\sigma}}(\ell_{i\tau})
= |w_{\sigma}(\ell_{i\tau})|^2$ is indeed real and positive
definite. The QMC simulations of KMH model at half-filling are therefore free of the sign problem.

\begin{figure}[tp]
\centering
\includegraphics[width=0.7\textwidth]{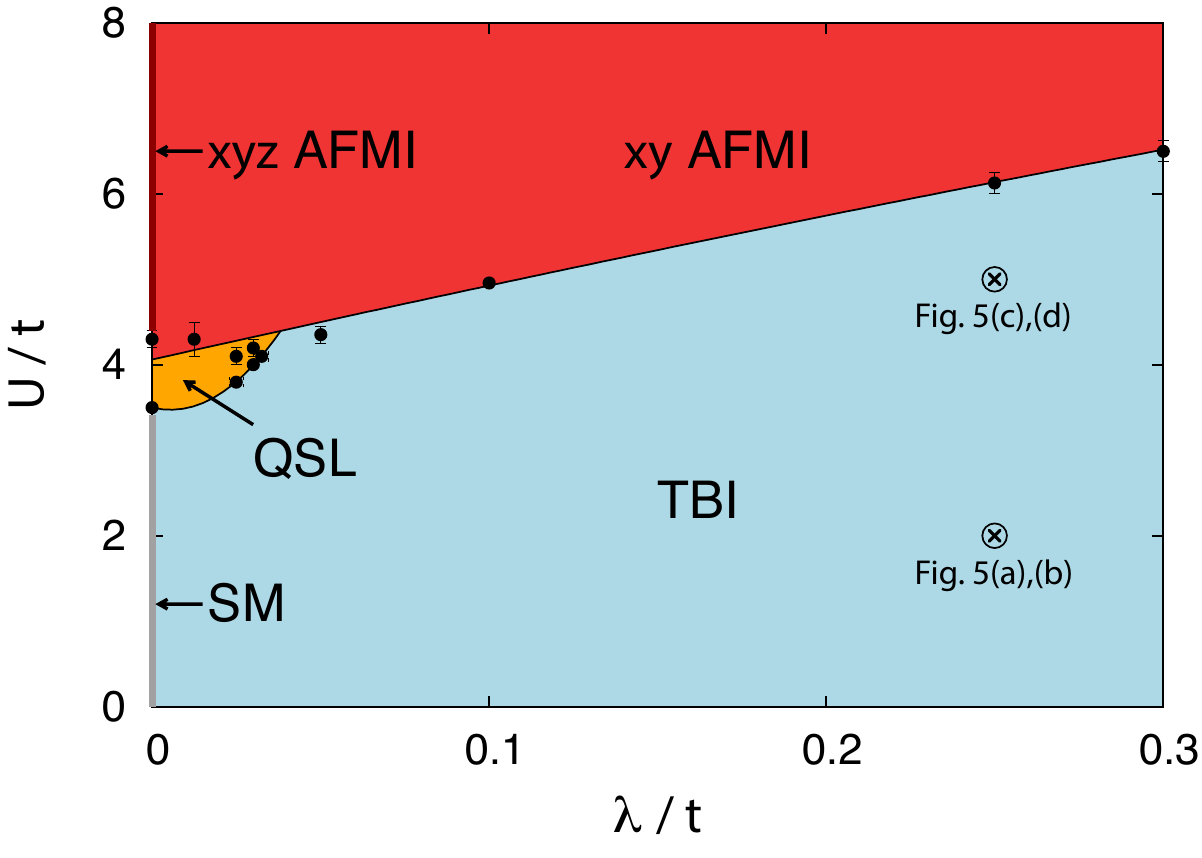}
\vspace*{8pt} \caption{Phase diagram of the Kane-Mele-Hubbard model
obtained from QMC simulations.\protect\cite{Lang2011,Hohenadler2012}
The phases are a quantum spin Hall insulator (TBI), a semimetal (SM
for $\lambda=0$), a quantum spin liquid (QSL), and an
antiferromagnetic Mott insulator (AFMI) with either Heisenberg (for
$\lambda=0$) or easy plane (for $\lambda\ne0$) order. Adapted and
reproduced with permission from Ref.~\onlinecite{Hohenadler2012}.
Copyright 2012 American Physical Society.} \label{fig:KMH_phasediagram}
\end{figure}

\begin{figure}[tp]
\centering
\includegraphics[width=\textwidth]{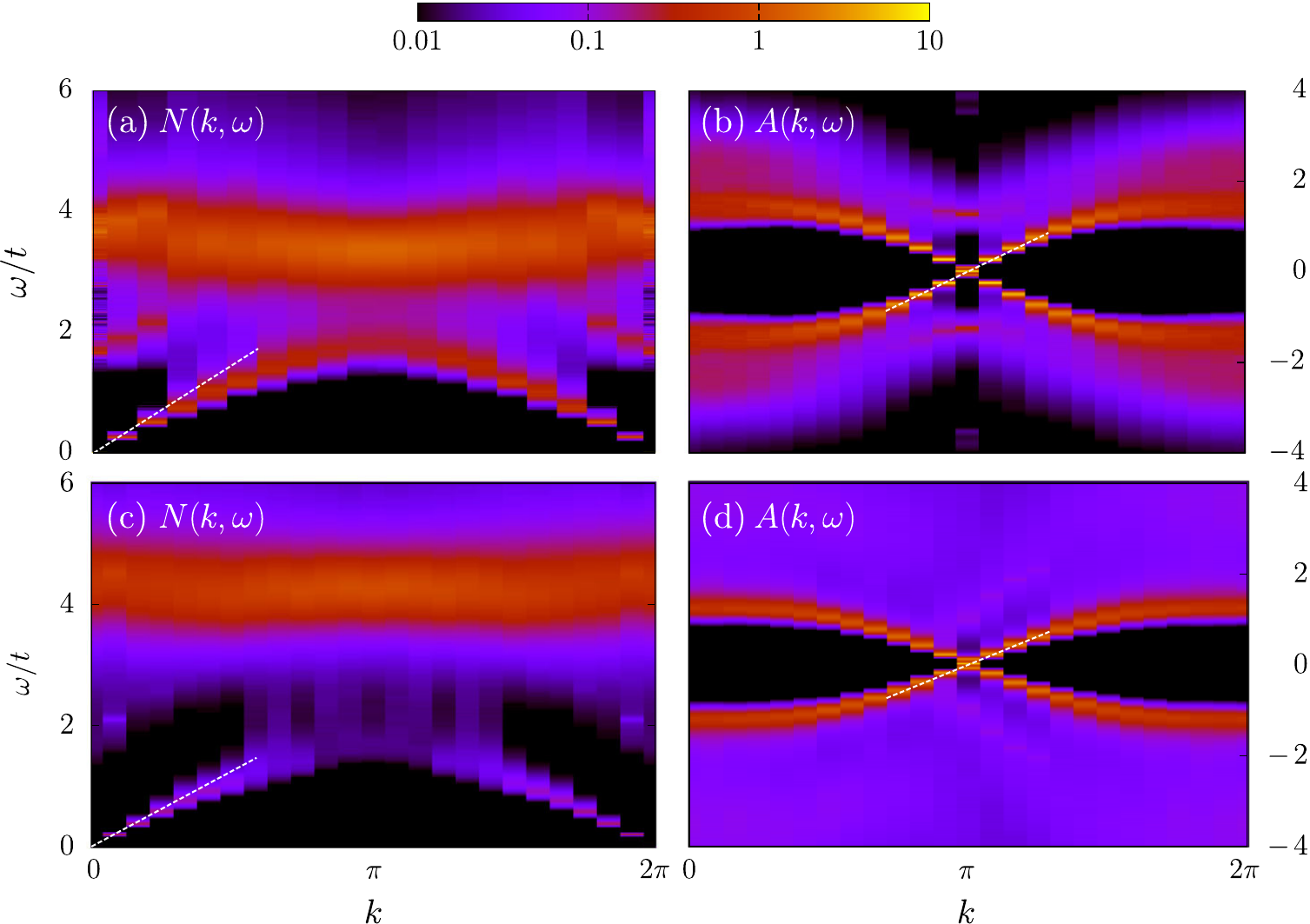}
\vspace*{8pt}
\caption{Dynamical spectra in the (a), (c) charge and (b), (d)
single-particle sectors, measured along the zigzag edge of a ribbon
geometry at ${\lambda/t=0.25}$ and ${U/t=2}$ (top panels), ${U/t=5}$
(bottom panels), respectively. Dotted lines show the excitation
velocities of the free system for comparison. Adapted and reproduced
with permission from Ref.~\onlinecite{Lang2011}. Copyright
2012 American Physical Society.} \label{fig:KMH_spectra}
\end{figure}

Without sign problem, the QMC method allows to efficiently measure
equal-time and time-displaced correlation functions, such as the
single-particle Green's functions\cite{Meng2010,Assaad1996,Feldbacher2001}
\begin{eqnarray}
   G_{\sigma}(\mathbf{k},\tau)=\langle \Psi_0 | c_{\mathbf{k}\sigma}(\tau)c^{\dag}_{\mathbf{k}\sigma}(0)|\Psi_0\rangle\;.
\end{eqnarray}
The single-particle gap $\Delta_\text{sp}(\mathbf{k})$ can be then
determined from the long imaginary time behavior of the
time displaced single-particle Green's function, i.e., ${G(\mathbf{k},\tau\to\infty)
\propto \exp(-\tau\Delta_\text{sp}(\mathbf{k}))}$. For the KM model at 
half-filling the relevant momenta are at the
Dirac points $K_1$ and $K_2$. The uniform
single-particle gap obtained from
${\sum_{\mathbf{k}}G(\mathbf{k},\tau\to\infty) \propto
\exp(-\tau\Delta_\text{u})}$ is used to describe the single-particle
gap, independently of a specific momentum. The gap for spin
excitations is obtained similarly from the imaginary-time displaced
spin-spin correlation function, for example in the antiferromagnetic
ordering (staggered) sector,
\begin{equation}
   S(\mathbf{k},\tau) = \langle\!\langle (\mathbf{S}_{A}(\mathbf{k},\tau)-\mathbf{S}_{B}(\mathbf{k},\tau))
      \cdot(\mathbf{S}_{A}(\mathbf{k},0)-\mathbf{S}_{B}(\mathbf{k},0)\rangle\!\rangle \;.
   \label{eq:dynamic_spinspincorr}
\end{equation}
The double brackets $\langle\!\langle \cdots\rangle\!\rangle$ denote
the cumulant of a correlation function of operators
${\langle\!\langle O_{1}O_{2}\rangle\!\rangle:=\langle
O_{1}O_{2}\rangle-\langle O_{1}\rangle\langle O_{2}\rangle}$. The
spin gap are obtained from
${S(\mathbf{k},\tau)\propto\exp(-\tau\Delta_{s}(\mathbf{k}))}$. 
Antiferromagnetic order in the honeycomb lattice corresponds to the
momentum at $\mathbf{k}=\Gamma$, hence, $\Delta_s=\Delta_s(\Gamma)$.
As for the static antiferromagnetic structure factor, it can be
obtained directly from the equal time (static) spin-spin correlation
function in the staggered sector at the momentum point $\Gamma$.
Note, that the intrinsic spin-orbit coupling term breaks
the SU(2) spin rotational invariance down to
the U(1) symmetry group, such that for
${\lambda>0}$ spontaneous spin symmetry breaking will occur in the
transversal spin
channel.\cite{Rachel2010,Lang2011,Zheng2011,Hohenadler2012} Hence it
is necessary to monitor $z$- and $xy$-spin correlations
independently.

Figure~\ref{fig:KMH_phasediagram} shows the phase diagram of KMH
model at half-filling obtained from QMC
simulations.\cite{Lang2011,Zheng2011,Hohenadler2012} Along the
$\lambda=0$ axis, where there is no spin-orbit coupling, the system
is a semimetal at small interaction $U/t$ with Dirac points shown in
Fig.~\ref{fig:kmtopologicalphasetransition}(a), and is an
antiferromagnetically order Mott insulator at large $U/t$ with
Heisenberg type order ($xyz$ AFMI). The existence of the phase in
the intermediate interaction strength (a possible quantum spin
liquid state) and the nature of the semi-metal to antiferromagnetic
insulator transition is under intensive
debate.\cite{Meng2010,Sorella2012,Assaad2013b} For any finite
$\lambda$, the system is in the QSH state, here named as topological
band insulator (TBI). At finite interaction $U/t>0$, the system
(indicated by blue region) is adiabatically connected to the
noninteracting case, e.g., the KM model. A stronger interaction
(i.e., at $\lambda=0.1t$, $U \gtrsim 4.5t$) will drive the TBI
through a continuous quantum phase transition into an
antiferromagnetic ordered Mott insulator (AFMI). The single-particle
gap remains open, but the spin gap closes. At finite
$\lambda$, the SU(2) spin symmetry is already broken
down to U(1) such that the magnetic order in the strong coupling
regime is in the $xy$ plane (easy plane) of spin space. The
transition from TBI to the $xy$ AFMI has been shown to be consistent
with the 3D XY universality class.\cite{Hohenadler2012}

To explore the correlation effect on the time-reversal symmetry
protected edge states, the authors in Ref.~\onlinecite{Lang2011}
studied the KMH model on the zigzag ribbon (as shown in the inset of
Fig.~\ref{f2}), and obtain the spectral information along the zigzag
edge of the ribbon. Figure~\ref{fig:KMH_spectra} shows
how the edge states evolve in the TBI phase under the increasing
influence of correlations. The panels show the single-particle
spectral function
\begin{equation}
   A(k,\omega) = \frac{1}{Z} \sum_{n,m,\sigma} \left({\rm e}^{-\beta E_n} + {\rm e}^{-\beta E_m} \right) |\langle m| c^{\dagger}_{k\sigma} |n\rangle |^2 \delta(E_m-E_n+\omega)\;,
\end{equation}
and the dynamic charge structure factor
\begin{equation}
   N(k,\omega) = \frac{1}{Z} \sum_{n,m} {\rm e}^{-\beta E_n} |\langle m| c^{\dagger}_{k\sigma}c_{k\sigma} |n\rangle |^2 \delta(E_m-E_0-\omega)\;,
\end{equation}
along the ribbon edge of an open system. At small interaction strength ${U/t=2}$
[panels (a) and (b)] signatures of the edge states can be clearly
seen below the bulk gap as linear mode around ${k=0}$. However, as
the interaction strength approaches the critical value, i.e.
${U/t=5}$ in panels (c) and (d), one observes a strong depletion of
spectral weight in the low-lying charge modes in (c), which leads to
reduction of the Drude weight. As the interaction strength ${U/t=5}$
is still below $U_c$ above which the transverse antiferromagnetic
order sets in, despite strong correlations, the single-particle
spectrum (d) still exhibits the helical edge states, which remain
essentially unaffected by the increased correlations.

\section{Detecting Topological Orders}

As discussed in Sec.~\ref{sect:kanemelemodel}, the quantity to
distinguish the QSH state from a trivial band insulator is the $Z_2$
invariant. Physically, the $Z_2$ topological invariant is associated
with the change in the time-reversal polarization when a magnetic flux
is threaded through a cylinder geometry varying
from $0$ to $h/2e$.\cite{Fu2006prb,Kane2007} Though this picture was
initiated in the noninteracting limit, with interaction one can
still observe similar behavior. In Sec.~\ref{sect:piflux}, we shall
show that, in the KM model the insertion of $\pi$-fluxes gives rise
to a Kramers doublets of spin-fluxon states.\cite{Assaad2013} We
then move our discussion to the evaluation of the $Z_2$ invariant.
Here, likewise, the construction of the topological invariant was
also initially defined in the noninteracting
limit.\cite{Kane05b,Fu2007prb,Fu2006prb,Kane2007} The $Z_2$
invariant, however, can be straightforwardly generalized to
interacting cases\cite{Wang2012prx,Wang2012a} and can be expressed
in terms of single-particle Green's functions. An overview of
to the $Z_2$ index and the parity behavior of the single-particle Green's 
function will be provided in Sec.~\ref{sect:Z2inv}. To illustrate the formalism and
demonstrate its power, in Sec.~\ref{sect:GDKMmodel} we
investigate the interacting topological phase transitions in two
descendants of the KMH model within QMC simulations, the generalized
KMH model\cite{Hung2013b,Hung2013} and the dimerized KMH model.
\cite{Lang2013} Moreover, we point out the limitation of the $Z_2$ topological
invariant approach in the QMC method. Sec.
\ref{sect:QMClimitation} will render an example which illustrates
the invariant's shortcoming to describe quantum phase
transitions which involve spontaneous symmetry breaking as a consequence from
collective excitations. Finally, in Sec.~\ref{sect:spinChern} we will discuss 
the evaluation of the spin Chern number from the sum over real-space derivatives 
of products of the eigenvectors of the zero-frequency Green's
functions.\cite{Hung2013b}

\subsection{$\pi$-flux Insertion}
\label{sect:piflux}

\begin{figure}[tp]
\centering
\includegraphics[width=\textwidth]{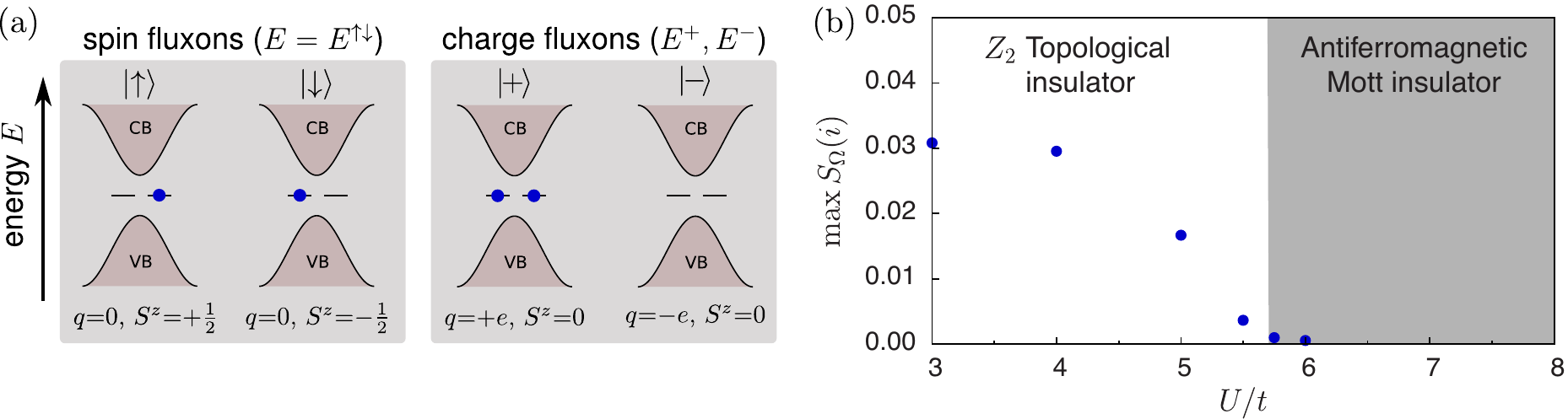}
\caption{ (a)
The four mid-gap fluxon states induced by a $\pi$-flux and their
associated charge $q$ and spin $Sz$, localized between valence bands
(VB) and conduction bands (CB) near the flux. (b) The maximum of the
site resolved integrated dynamical structure factor $S_{\Omega}(i)$
across the phase transition into the $xy$ antiferromagnetic regime,
indicating the absence of spin fluxons in the magnetic phase.
Adapted and reproduced with permission from
Ref.~\onlinecite{Assaad2013}. Copyright 2013 American Physical
Society.} \label{fig:piflux}
\end{figure}

The authors of Refs. \onlinecite{Ran2008} and \onlinecite{Qi2008}
have shown, that on a lattice with periodic boundaries, $\pi$-fluxes
can be inserted in pairs, threading selected plaquettes of the
lattice. In the topological phase, each $\pi$-flux gives rise to
four fluxon states near the corresponding flux-threaded hexagons.
The states correspond to the spin fluxons ${|\uparrow\rangle}$,
${|\downarrow\rangle}$, forming a Kramers pair related by time
reversal symmetry, and charge fluxons ${|+\rangle}$, ${|-\rangle}$,
related by particle-hole symmetry as illustrated in
Fig.~\ref{fig:piflux}(a). As a consequence of the bulk gap, these
states are exponentially localized around the flux-threaded
plaquettes and energetically lie inside the bulk band gap.

Assaad {\it et al.}\cite{Assaad2013} have successfully shown, that
two maximally separated $\pi$-fluxes can be used to probe the
correlated quantum spin Hall state for its topological properties.
The $\pi$-flux pairs introduce edges in the bulk system around which
these spin fluxons manifest. Spin fluxons can then be detected by
calculating the lattice-site-resolved, dynamical spin-structure
factor at zero temperature, defined as
\begin{equation}
   S(i,\omega) = \pi \sum_n |\langle n|S_i^z|0\rangle|^2\, \delta(E_n-E_0-\omega)\;.
\end{equation}
Here, $S(i,\omega)$ corresponds to the spectrum of spin excitations
at lattice site $i$, ${H|n\rangle =E_n|n\rangle}$ defines the
excitation energies and $|0\rangle$ denotes the ground state. The
dynamical spin-structure factor picks up the spin-fluxon states.
Integration of $S(i,\omega)$ up to an energy scale $\Omega$ well
below the charge gap ${\Omega\ll\Delta_{c}\approx 2
\Delta_\text{sp}}$, allows to account for all spin-fluxon excitations.
This yields the site resolved integrated dynamical structure factor
${S_{\Omega}(i)=\int_0^{\Omega}d\omega\, S(i,\Omega)}$, which can be
used to identify the presence of spin-fluxons in the topological
insulator, or lack thereof in topologically trivially ordered
phases. The dependence of ${S_{\Omega}(i)}$ is demonstrated in
Ref.~\onlinecite{Assaad2013} for the KMH model across the magnetic
quantum phase transition from TBI to the $xy$ AFMI at ${\lambda/t = 0.2}$. In
Fig.~\ref{fig:piflux}(b) the maximum of $S_{\Omega}(i)$ is plotted
as a function of the interaction strength ${U/t}$. The observable
acquires finite values  in the topological-insulator phase, and a
strong drop is observed on approaching the critical point
${U_c/t\approx 5.7}$, before it vanishes in the magnetically ordered
phase. The spin-fluxon signal can be used in quantum Monte 
Carlo simulations as a general tool to distinguish topological and 
nontopological phases, although the need for the continuation to real 
frequencies can make its use impractical, or
result in lack of accuracy.

In addition to the integrated dynamical structure factor at ${T=0}$, at
finite temperature spin fluxons created by $\pi$-fluxes give rise to
a characteristic Curie law in the spin susceptibility, which can be
used to identify topological properties in finite temperature
quantum Monte Carlo simulations. At low temperatures, the spin
susceptibility ${\chi = \beta (\langle S^2_z\rangle-\langle
S_z\rangle^2)}$ then follows the form ${\chi \sim 2/k_\text{B}T}$,
or ${1/k_\text{B}T}$ per $\pi$-flux. For details on further results
of $\pi$-fluxes and the interactions between the induced spin
fluxons we refer the reader to Ref.~\onlinecite{Assaad2013}.

\subsection{The $Z_2$ Topological Invariant}
\label{sect:Z2inv}

\subsubsection{The Parity Invariant and the Zero-frequency Green's Function}
\label{sect:Z2formalism}

The QSH (topological insulator) state is identified by the $Z_2$
invariant $\nu=1$. In Refs. \onlinecite{Fu2007prb,Fu2006prb}, the
$Z_2$ invariant is defined as
\begin{eqnarray}
    (-1)^{\nu}=\prod_{\gbf{\kappa}}\delta_{\gbf{\kappa}}\;,
\end{eqnarray}
where $\gbf{\kappa}$ denotes the time-reversal invariant momenta
(TRIM) of the Brillouin zone. In two dimensional QSH states, there are four TRIM
points, whereas in three dimensional topological insulators there are eight. For the
honeycomb lattice, the TRIM have been introduced in
Fig.~\ref{fig:KMmodel}(b). The value of the number
$\delta_{\gbf{\kappa}}$ is evaluated according to
Ref.~\onlinecite{Fu2006prb} as
\begin{eqnarray}
   \delta_{\gbf{\kappa}}=\frac{\sqrt{\det{[w(\gbf{\kappa})]}}}{\textrm{Pf}[w(\gbf{\kappa})]}=\pm 1. \label{eq:pfifian}
\end{eqnarray}
Here, $w(\gbf{\kappa})$ is an antisymmetric matrix with the elements
defined by ${[w(\gbf{\kappa})]_{mn}=\langle u_{m
-\gbf{\kappa}}|\mathcal{T}| u_{n\gbf{\kappa} } \rangle}$ and $m$, $n$
stand for the band indices. $\mathcal{T}$ denotes the time-reversal
operator, $\mathcal{T}^2=-1$ for spin-1/2, and
$|u_{i{\gbf{\kappa}}}\rangle$ is the Bloch state of the
noninteracting Hamiltonian of Eq.~(\ref{eq:kmhamiltonianmatrix}),
i.e., $H(\gbf{\kappa})|
u_{i{\gbf{\kappa}}}\rangle=E_i(\gbf{\kappa})|
u_{i{\gbf{\kappa}}}\rangle $. Pf$[w]$ denotes the Pfaffican function
of the matrix $w(\gbf{\kappa})$, with $\det{[w]}=\textrm{Pf}[w]^2$.
Note that due to the presence of the square root, the sign of
$\delta_{\gbf{\kappa}}$ is ambiguous. $|u_{i{\gbf{\kappa}}}\rangle$
should be chosen continuously in the Brillouin zone, so that
$\sqrt{\det{[w]}}$ is defined globally.\cite{Fu2007prb}

In the presence of inversion symmetry, which is the case for the KM
model without Rashba coupling, Eq.~(\ref{eq:pfifian}) can be simply
evaluated as\cite{Fu2007prb}
\begin{eqnarray}
   \delta_{\gbf{\kappa}}=\prod_{m}\xi_{2m}(\gbf{\kappa})\;,
   \label{eq:noninteractingparity}
\end{eqnarray}
where $\xi_{2m}(\gbf{\kappa})$ denotes the parity eigenvalue of the 
$2m$-th occupied band at momentum $\gbf{\kappa}$. Here we used $2m$ 
to indicate that there exists the Kramers degenerate pair 
$\xi_{2m}(\gbf{\kappa})$ and $\xi_{2m-1}(\gbf{\kappa})$ with the 
same parity value. For sustained inversion symmetry, the Bloch states 
are also eignestates of the inversion operators, so $\xi_{2m}(\gbf{\kappa})=\pm 1$. 
Rather than Eq.~(\ref{eq:pfifian}), Eq.~(\ref{eq:noninteractingparity}) 
is obviously more practical and simpler to evaluate the $Z_2$ invariant.

Equations (\ref{eq:pfifian}) and (\ref{eq:noninteractingparity}) are 
only suitable in the noninteracting limit. In the presence of interaction, 
the Bloch states are no longer well-defined. However, it has been shown 
that for interacting topological insulators the topological order 
parameters can be expressed in terms of Green's functions defined in 
the extended frequency-momentum space\cite{Wang2010}
\begin{eqnarray}
   P_3 = \frac{\pi}{6}\int^1_0 \!\! du \int \!\! \frac{d^4k}{(2\pi)^4}\textrm{Tr}[\epsilon^{\mu \nu \rho \sigma} G \partial_{\mu} G^{-1}\, G
   \partial_{\nu}G^{-1}\, G\partial_{\rho} G^{-1}\, G\partial_{\sigma}G^{-1}\, G\partial_u G^{-1}]\;,
   \label{eq:topomagneteectrcoef}
\end{eqnarray}
in which $G=G(k,u)$, $k=(\omega,\mathbf{k})$. The momentum
$\mathbf{k}$ are integrated over the Brillouin zone and the
frequency $\omega$ is integrated over $(-\infty,+\infty)$.
The extra dimension $u$ in $G(k,u)$ has the following meaning: ${u=0}$ 
corresponds to the Green's function for interacting topological insulator, 
${u=1}$ corresponds to the Green's function of a trivial insulator. 
Values of ${u\in(0,1)}$ smoothly connect the two limits.
Equation ~(\ref{eq:topomagneteectrcoef})
can be interpreted as the physical response of an
insulator in the topological field theory.\cite{qithuges2008prb}
This formula however involves the full
frequency-momentum space integral and an extra-dimension $u$ where
one extends the topological insulator to a topologically trivial
insulator. Thus it is apparently not practical
to implement this formula in numerical simulations. Fortunately, the
authors of Ref.~\onlinecite{Wang2012prx} showed that
$G^{\prime}({\mathbf k},\I\omega)=1/(\I\omega +G^{-1}({\mathbf
k},0))$ is topologically equivalent to the nonzero frequency Green's
function $G({\mathbf k},\I\omega)$, and hence this topological order
parameter can be simply expressed in terms of the Green's function
at zero frequency.\cite{Wang2012prx,Wang2012a,Wang2012c}
$G^{-1}({\mathbf k},0)$ is further interpreted as the topological
Hamiltonian, which contains all necessary information of the
existence of surface states.\cite{Wang2013JPC} This greatly
simplifies numerical and analytical calculations.

The procedure to obtain the topological invariant from the
zero-frequency Green's function is hence described as follows:
Diagonalize the inverse zero-frequency Green's function at the TRIM
\begin{eqnarray}
   G^{-1}(\gbf{\kappa},0)|\mu_n(\gbf{\kappa},0) \rangle=\mu_n |\mu_n(\gbf{\kappa},0)\rangle \;,
\end{eqnarray}
to acquire the state $|\mu_{n}(\gbf{\kappa},0) \rangle$, and choose
the eigenvectors associated with positive eigenvalues ($\mu_n>0$,
denoting the generalization of {it occupied bands} and are
called right-zero, or R-zero).\cite{Wang2012prx} The R-zeroes span the
R-space at each $\gbf{\kappa}$. In analogy to
Eqs.~(\ref{eq:pfifian}) and (\ref{eq:noninteractingparity}), for an
interacting topological insulator with inversion symmetry we have
\begin{eqnarray}
   (-1)^{\nu}=\prod_{\gbf{\kappa}}\frac{\sqrt{\det{[W(\gbf{\kappa})]}}}{\textrm{Pf}[W(\gbf{\kappa})]}\;,
\end{eqnarray}
The matrix elements $[W(\gbf{\kappa})]_{mn}=\langle
\mu_{m}(-\gbf{\kappa},0)|\mathcal{T}|\mu_{n}(\gbf{\kappa},0)
\rangle$. In inversion symmetric systems $|\mu_{n}(\gbf{\kappa},0)
\rangle$ is the simultaneous eigenvector of $G(\gbf{\kappa},0)$ and
the parity (or inversion) operator $\mathcal{P}$
\begin{eqnarray}
   P|\mu_n(\gbf{\kappa},0) \rangle= \eta_{\mu}(\gbf{\kappa},0) |\mu_n(\gbf{\kappa},0)\rangle\;.
\label{eq:parity}
\end{eqnarray}
This means the evaluation of the $Z_2$ (parity) invariant only relies 
on the structure of the zero-frequency Green's function $G(\gbf{\kappa},0)$, 
which can be readily computed within auxiliary field QMC simulations.

In order to compute the Matsubara Green's function from the imaginary-time 
displaced Green's function, and continue in particular to zero frequency, 
let us first consider the system at a finite temperature $T=1/\beta$. The 
finite temperature, imaginary time Green's function $G_\sigma(\mathbf{k},\tau;\beta)$ 
with $\tau\in[0,\beta]$, is a $2\times 2$ matrix expressed as
\begin{equation}
   [G_\sigma(\mathbf{k},\tau;\beta)]_{jl}=-\langle c_{\mathbf{k}\sigma j}(\tau)\, c^\dagger_{\mathbf{k}\sigma l}(0) \rangle_\beta \;,
\end{equation}
where ${j,l=}$A, B are sublattice indices of the honeycomb lattice.
To be represented in Matsubara-frequencies ${\omega_n=(2n+1)\pi/\beta}$, one needs to perform
the Fourier transformation
\begin{equation}
   G_{\sigma}(\mathbf{k},\I\omega_n;\beta)=\int_0^\beta G_{\sigma}(\mathbf{k},\tau;\beta)\, \E^{\I \omega_n \tau} d\tau \;.
\end{equation}
The particle hole symmetry of the half-filled KMH model, i.e., under
the transformation ${c^\dagger_{\mathbf{k}\sigma j}\rightarrow
d_{\mathbf{k}\sigma j} = (-1)^j c^\dagger_{-\mathbf{k}\sigma j}}$
in each spin sector together with inversion symmetry leads to the
following conditions on $G_{\sigma}(\mathbf{k},\tau;\beta)$: For
equal sublattices,
$[G_\sigma(\mathbf{k},\tau;\beta)]_{jj}=[G_\sigma(-\mathbf{k},\beta-\tau;\beta)]_{jj}$,
while, for $j\neq l$,
${[G_\sigma(\mathbf{k},\tau;\beta)]_{jl}=-[G_\sigma(-\mathbf{k},\beta-\tau;\beta)]_{jl}}$.
We thus obtain for the diagonal elements of the Green's function at
one of the TRIM $\gbf{\kappa}$ points
\begin{equation}
   [G_{\sigma}(\gbf{\kappa},\I \omega_n;\beta)]_{jj}=2\,\I\int_0^{\beta/2} [G_\sigma(\gbf{\kappa},\tau;\beta)]_{jj} \sin(\omega_n \tau)\, d\tau \;,
\end{equation}
and, for $j\neq l$,
\begin{equation}
   [G_{\sigma}(\gbf{\kappa},\I \omega_n;\beta)]_{jl}=
      2 \int_0^{\beta/2} [G_\sigma(\gbf{\kappa},\tau;\beta)]_{jl} \cos(\omega_n \tau)\, d\tau \;.
\end{equation}
Now, the limit $\beta\rightarrow\infty$ can be taken properly: From the 
projective QMC simulations, we obtain the ground state Green's function 
$G_\sigma(\gbf{\kappa},\tau)=\lim_{\beta\rightarrow\infty}G_\sigma(\gbf{\kappa},\tau;\beta)$, 
and then perform the above integrals with a sufficiently large cutoff 
$\beta\rightarrow\theta$, set e.g. by the imaginary time evolution length 
of the Green's function $\theta$ employed in the QMC simulations (we 
usually use  ${\theta=20/t}$). Note, that one cannot simply take the 
limit ${\I\omega_n\rightarrow 0}$ before accounting for the (anti)symmetry 
conditions on the imaginary time Green's functions. This would lead to 
wrong results, as exemplified below. After (anti)symmetrization, the limit 
$\I\omega_n\rightarrow 0$ can be performed with the ${T=0}$ Green's 
functions, so that
\begin{equation}
   [G_{\sigma}(\gbf{\kappa},\omega=0)]_{jj}=0 \;,
   \label{Gstructure1}
\end{equation}
and, for $j\neq l$,
\begin{equation}\label{eq:gomegajl}
   [G_{\sigma}(\gbf{\kappa},\omega=0)]_{jl}= 2\int_0^{\theta/2} [G_\sigma(\gbf{\kappa},\tau)]_{jl}\, d\tau \;.
   \label{Gstructure2}
\end{equation}
Note, that this structure of the Green's function is a direct consequence of the common eigenvector system shared by $G_{\uparrow}(\gbf{\kappa},0)=G_{\downarrow}(\gbf{\kappa},0)$, and $G(\gbf{\kappa},0)$ with $\mathcal{P}$, such that the one has the relation
\begin{eqnarray}
    G(\gbf{\kappa},0)=\alpha_{\gbf{\kappa}} \gbf{\sigma}^x \;.
    \label{eq:gamma}
\end{eqnarray}
Hence, within the QMC simulations, one merely needs to measure the
off-diagonal part of the Green's function explicitly. To illustrate
the above point, consider the exact ${T=0}$
imaginary-time Green's function
\begin{equation}
   G_{\sigma}(\gbf{\kappa},\tau)= -\frac{1}{2}\, \E^{-|g_{\gbf{\kappa}}|\tau}\left(
   \begin{matrix}
      1 & -1\\
     -1 &  1
   \end{matrix}
\right)\;.
\end{equation}
If calculated naively, via ${\int_0^\infty G_{\sigma}(\tau,\Gamma)
\,d\tau}$, one would (wrongly) obtain a finite value of
$[G_{\sigma}(\omega=0,\Gamma)]_{jj}$ instead of the actual value
(i.e. zero).

The KMH model has the explicit $S^{z}$ conservation of the
Hamiltonian (spin independent motion). Thus the single-particle
Green's function is block-diagonal in spin-space ${G({\mathbf
k},0)=G_{\uparrow}({\mathbf k},0) \oplus G_{\downarrow}({\mathbf
k},0)}$, and the procedure of calculating
$\eta_{\mu}(\gbf{\kappa},0)$ in Eq. (\ref{eq:parity}) can be
restricted to $G_{\uparrow}({\mathbf k},0)$, or
${G_{\downarrow}({\mathbf k},0)}$. The Green's function
${G_{\sigma}(\mathbf{k},\I\omega=0)}$ is a ${2\times2}$ matrix in the
A/B-sublattice basis. In the spinor convention
${\Psi^{\dag}=(c^{\dag}_{{\rm A},\uparrow} c^{\dag}_{{\rm B},\uparrow} \
c^{\dag}_{{\rm A},\downarrow} c^{\dag}_{{\rm B},\downarrow})}$, the parity
operator of the honeycomb lattice is defined as ${\mathcal{P}=
\mathbf{1} \otimes \gbf{\sigma}^{x}}$,\cite{Fu2007prb} which
interchanges A and B sublattices. Note that since $G$ and
${G^{-1}}$ have the same eigenvectors, we can directly diagonalize
${G_{\sigma}({\mathbf{k}},0)
=[-H_{\mathbf{k}}-\Sigma({\mathbf{k}},0)]^{-1}}$ instead of the
inverse Green's at the TRIM
\begin{eqnarray}
   G_{\sigma}({\gbf{\kappa}},0)|\tilde{\mu}_n(\gbf{\kappa},0) \rangle=\tilde{\mu}_n |\tilde{\mu}_n(\gbf{\kappa},0)\rangle\;,
\end{eqnarray}
and then choose the R-zero eigenvectors (${\tilde{\mu}_n=1/\mu_n>0}$)
to evaluate the corresponding parity
${\tilde{\eta}_{\mu}(\gbf{\kappa})=\langle
\tilde{\mu}_n(\gbf{\kappa},0) |\mathcal{P}|
\tilde{\mu}_n(\gbf{\kappa},0) \rangle}$. For the honeycomb lattice,
$n=1$ at half filling, thus each ${\mathbf k}$ has one R-zero and we
can simplify the notation
${\tilde{\eta}_{\mu}(\gbf{\kappa})=\tilde{\eta}_{\gbf{\kappa}}}$. Also
at these TRIM, Kramers degenerate partners share the same parity
eigenvalues and one can restrict the procedure to one spin sector,
say $G_{\uparrow}(\gbf{\kappa},0)$, to compute
$\tilde{\eta}_{\gbf{\kappa}}$, and then
\begin{equation}
   (-1)^{\nu}=\prod_{\gbf{\kappa}} \tilde{\eta}_{\gbf{\kappa}}\;. \label{eq:z2invariant}
\end{equation}
The value of $\nu=0$ denotes a trivial insulator,
whereas $\nu=1$ indicates a $Z_2$ topological insulator. In the
following section, we will present two example cases to identify the
interacting QSH state using Eq.~(\ref{eq:z2invariant}).

We will show in Sec.~\ref{sect:GDKMmodel} that in addition to the
$Z_2$ invariant, the proportional coefficient
$\alpha_{\gbf{\kappa}}$ in Eq.~(\ref{eq:gamma}) can also be used to
characterize the $Z_2$ topological insulator/trivial insulator phase
transition and even is more sensitive than $\nu$ numerically: at the
topological phase transition, where the bulk gap closes at the TRIM,
the zero-frequency single-particle Green's function is divergent on
the poles and $\alpha_{\gbf{\kappa}}$ flips the sign beyond the
transition.\cite{Gurarie2011} Here we want to point out, that while
in the KM model ($U=0$), ${\tilde{\eta}_{\gbf{\kappa}}=\pm1}$, in the
cases of finite $U$ and for interacting Green's functions ${\langle
\tilde{\eta}_{\gbf{\kappa}} \rangle=\pm1}$ is not guaranteed in a
single QMC measurement (cf. the supplemental material of
Ref. \onlinecite{Hung2013}). The well-defined parity invariant is
recovered only by acquiring sufficient statistics within the QMC
simulations. In the following we present cases, where this approach 
works, or breaks down respectively, and discuss the limitations of the 
use of the $Z_2$ invariant in simulations.

\subsubsection{Topological Phase Transitions in the Generalized and the Dimerized Kane-Mele-Hubbard Models}
\label{sect:GDKMmodel}

In this subsection, we will present two example case studies of
calculating the $Z_2$ parity invariant and the spin Chern number
within the QMC method. The models considered are descendants of the
KMH model, which we called the generalized KMH
model\cite{Hung2013b,Hung2013} and the dimerized KMH
model.\cite{Lang2013} Both of the models characterize a
$Z_2$-topological insulator to trivial-insulator phase transition as
a function of the tight-binding parameters.
Following Sec.~\ref{sect:QMCforKMH}, at half-filling both systems
are particle-hole and time-reversal symmetric. Therefore the QMC
simulations in these models are sign-free and we can accurately
determine the topological phase boundary at different values of $U$
beyond the mean-field level.

The main results are that, in the generalized KMH model, increasing
$U$ stabilizes the $Z_2$ topological insulator phase, whereas
the correlation effects in the dimerized KMH model destabilizes the
$Z_2$ topological insulator phase. In both cases the $Z_2$ invariant
proves to be a practical tool to determine the loss of the
topological nature as the systems are tuned into trivial band
insulators. Similar to the KMH model in Fig. \ref{fig:KMH_phasediagram}, 
the onsite Hubbard interaction in the descendant KMH models in the
thermodynamic limit will also induce spontaneous planar
antiferromagnetic order which breaks inversion
symmetry, such that Eq.~(\ref{eq:z2invariant}) is no longer
well-defined. Across the transition into the antiferromagnetic state
on finite size lattices the $Z_2$ invariant however fails to
reflect the change of the topological nature of the phases. The
recovery of its validity in the thermodynamic limit is not obvious
from the results obtained for increasing system sizes.

\vspace{2em}\noindent\textit{Generalized Kane-Mele-Hubbard model\vspace{1em}}\

\begin{figure}[tp]
   \centering
\includegraphics[width=\textwidth]{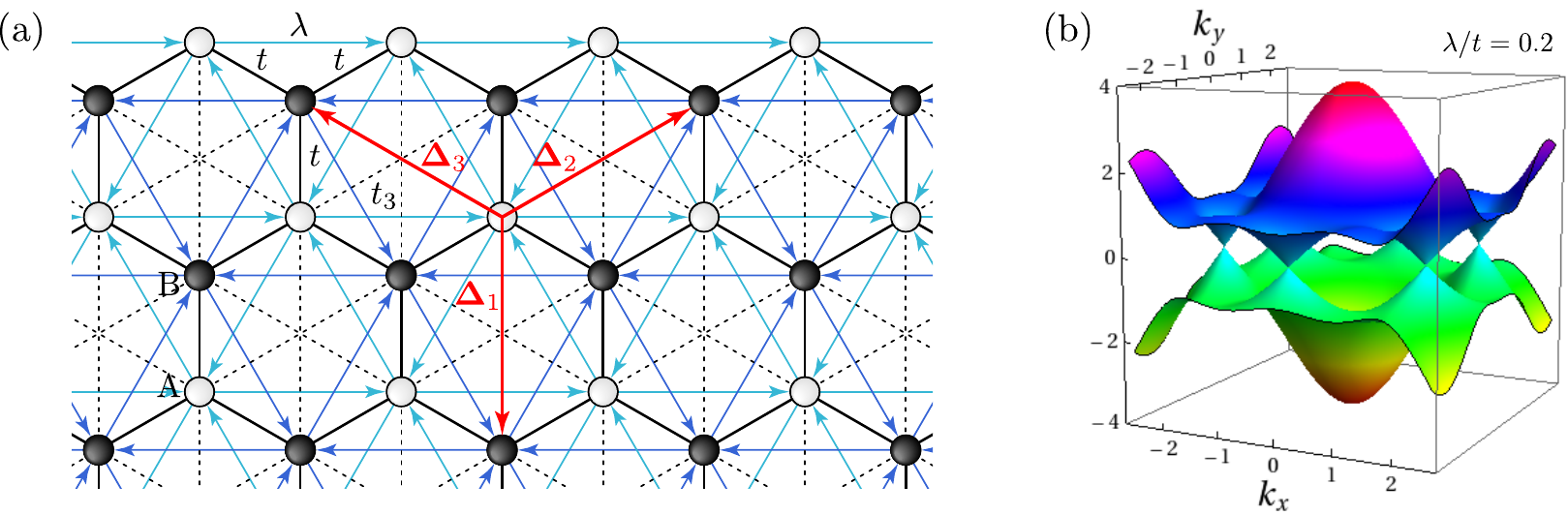} 
   \caption{(a)
The lattice geometry of the GKM model. The vectors
${\gbf{\Delta}}_{1,2,3}$ denote the directions of the third-nearest-neighbor hopping.
(b) The band structure of the GKM model at $t_3=\frac{1}{3}t$. The
band gap closes at the TRIM $M_{1,2,3}$ instead of at the Dirac
points $K_{1,2}$.}
   \label{fig:gkmtopologicalphasetransition}
\end{figure}

\noindent The generalized Kane-Mele model (GKM) is considered on the
KM lattice with real-valued third-neighbor hopping $t_{3}$
\begin{equation}
   H_\text{GKM}  =
      - t\sum_{\langle i,j\rangle,\sigma}c_{i\sigma}^{\dagger}c_{j\sigma}
      + \I\,\lambda\sum_{\langle\!\langle i,j \rangle\!\rangle, \alpha\beta} \nu_{ij}\,c^{\dagger}_{i\alpha}\sigma^{z}_{\alpha\beta}c_{j\beta}
      - t_{3}\sum_{\langle\!\langle\!\langle i,j\rangle\!\rangle\!\rangle,\sigma}c_{i\sigma}^{\dagger}c_{j\sigma}\;.
\label{eq:GKMHam}
\end{equation}
$\langle\!\langle\!\langle \cdots \rangle\!\rangle\!\rangle$ sums over 
third-nearest-neighbor hoppings described by vectors: ${\gbf{\Delta}_1=(0,-2)}$, 
${\gbf{\Delta}_2=(\sqrt{3},1)}$ and ${\gbf{\Delta}_3=(-\sqrt{3},1)}$, with 
the hopping amplitude of $t_3$ connecting $A$ and $B$ sublattices as indicated 
in Fig.~\ref{fig:gkmtopologicalphasetransition}(a). At $t_3=0$, the GKM model 
reduced to the KM model, and thus it is a topological QSH state.
In momentum space, the GKM model can be recast as $H_\text{GKM}=\sum_{\mathbf{k}}\Phi^{\dagger}_{\mathbf{k}}H(\mathbf{k})\Phi_{\mathbf{k}}$, with
\begin{equation}
    H(\mathbf{k}) = \left( \begin{array}{cccc}
   \gamma_{\mathbf{k}} & -f_{\mathbf{k}} & 0 & 0 \\
   -f^{\ast}_{\mathbf{k}} & -\gamma_{\mathbf{k}} & 0 & 0 \\
   0 & 0 & -\gamma_{\mathbf{k}} & -f_{\mathbf{k}} \\
   0 & 0 & -f^{\ast}_{\mathbf{k}} & \gamma_{\mathbf{k}} \end{array}
   \right)\;. \label{eq:gkmhamiltonianmatrix}
\end{equation}
The off-diagonal term $f_{\mathbf{k}}$ is given by $f_{\mathbf{k}}=
g_{\mathbf{k}}+t_3\sum^{3}_{i=1}\E^{\I\mathbf{k}\cdot\gbf{\Delta}_{i}}$, 
where $g_{\mathbf{k}}$ comes from the KM model in 
Eq.~(\ref{eq:kmhamiltonianmatrix}). Since the $t_3$ hopping does not 
break the time-reversal symmetry, the GKM Hamiltonian is still 
time-reversal symmetric. The dispersion of the GKM model is given by 
$\varepsilon_\text{GKM}(\mathbf{k})=\pm\sqrt{|f_{\mathbf{k}}|^{2}+\gamma^{2}_{\mathbf{k}}}$.

Beginning from $t_3=0$ and then moving to larger $t_3$, the GKM
model remains gapped, until at $t_3=\frac{1}{3}t$, the gap
collapses. This indicates that a topological phase transition occurs
at $t^c_3=\frac{1}{3}t$, and the regime of $0 \le t_3 < t^c_3$ is a
QSH state since it adiabatically connects to the $t_3=0$ case. We
have confirmed that the system in this regimes has $\nu=1$. The band
structure  of Eq.~(\ref{eq:gkmhamiltonianmatrix})  at $t^c_3$ is
shown in Fig.~\ref{fig:gkmtopologicalphasetransition} (b), where the
system exhibits three gapless modes located at the three TRIM
$M_{1,2}$ and $M_3$, rather than the Dirac points $K_{1,2}$.  On the
other hand, as $t_{3} > \frac{1}{3}t$, the band gap opens again, and
it is identified as $\nu=0$, a trivial insulator. At the
noninteracting level, the value of $t^{c}_3$ is independent of
$\lambda$.

\begin{figure}[tp]
\centering
\includegraphics[width=0.9\textwidth]{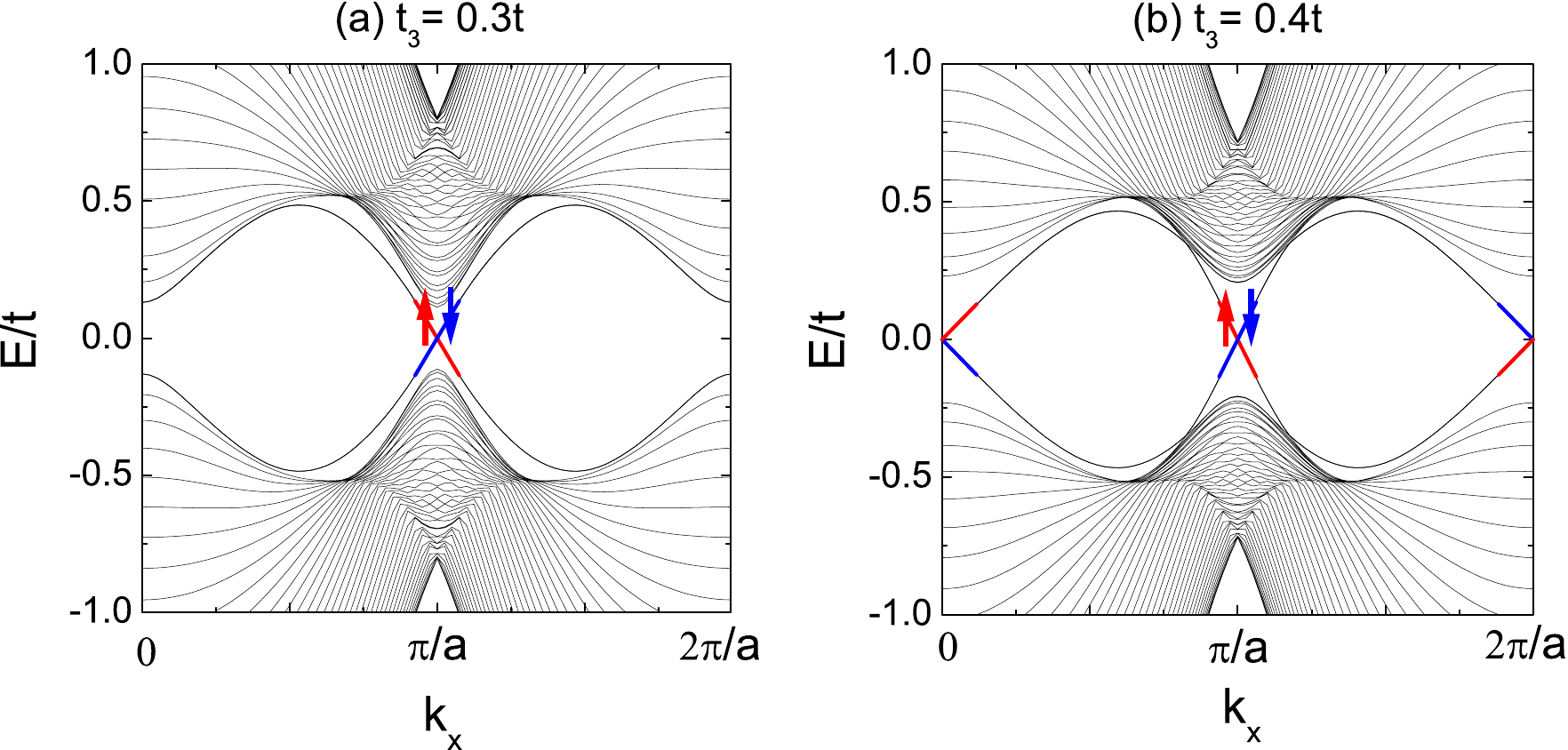}
   \caption{The edge spectra for the noninteracting GKM model at
$\lambda=0.1t$ and (a) $t_{3N}=0.3t$, a $Z_2$ topological
insulator and (b) $t_{3N}=0.4t$, a trivial insulator. The system is
considered as a zigzag ribbon geometry where an open boundary
condition is used along the zigzag direction.}
   \label{fig:GKMedgestate}
\end{figure}

To further understand the discrepancy between the $Z_2$ topological
band insulator phase and a trivial insulator phase, one can study
the edge modes with a zigzag ribbon geometry. Figures
\ref{fig:GKMedgestate}(a),(b) reveal different behavior of the edge
spectra for topologically nontrivial and trivial cases at
$\lambda=0.1t$. For the topological insulator phase at $t_{3}=0.3t$,
Fig.~\ref{fig:GKMedgestate}(a) shows an odd number of helical edge
states within the band gap, crossing at the time-reversal invariant
point $k_x=\pi$.
At $t_{3}=0.4t$ however, Fig.~\ref{fig:GKMedgestate}(b) shows an
even number of helical modes at $k_x=0$ and $k_x=\pi$, and thus from
the $Z_2$ perspective it is a topologically trivial state (the edge
states are not topologically protected). However, note that the two
helical states imply that the spin Chern number (discussed in Sec.~\ref{sect:spinChern}) $|C_{\sigma}|=2$,
and each spin sample is a $C=2$ IQH state. From the tight-binding
calculation, we determine that $C_{\sigma}=\pm1$ at $t_3 <
\frac{1}{3}t$, whereas $C_{\sigma}=\mp 2$ at $t_3 > \frac{1}{3}t$
for $\sigma={\uparrow,\downarrow}$. This corresponds to the
observation that the bulk band gap closes at three TRIM in
Fig.~\ref{fig:gkmtopologicalphasetransition}(b) ($|\Delta
C_{\sigma}|=3$).\cite{Hung2013b}

Next we augment the GKM model with an onsite Coulomb interaction.
This generalized KMH model is the GKM model plus an onsite Hubbard 
interaction, i.e., $H=H_\text{GKM}+H_U$ with ${H_U  = \frac{U}{2} \sum_{i} (n_{i}-1)^2}$. 
By sign-free QMC simulations, we can demonstrate how to identify a 
correlated topological insulator phase and a trivial insulating state 
with the single-particle Green functions and the $Z_2$ index. The 
simulations are performed using an imaginary time step ${\Delta \tau t=0.05}$ 
and an inverse temperature $\Theta t=40$. All the results use the periodic 
boundary conditions and the number of sites is ${N=2\times L^2}$, where 
$L$ is the linear system size.

\begin{figure}[tp]
   \begin{center}
      \begin{minipage}[t]{0.3\textwidth}\centering
         \includegraphics[width=\textwidth]{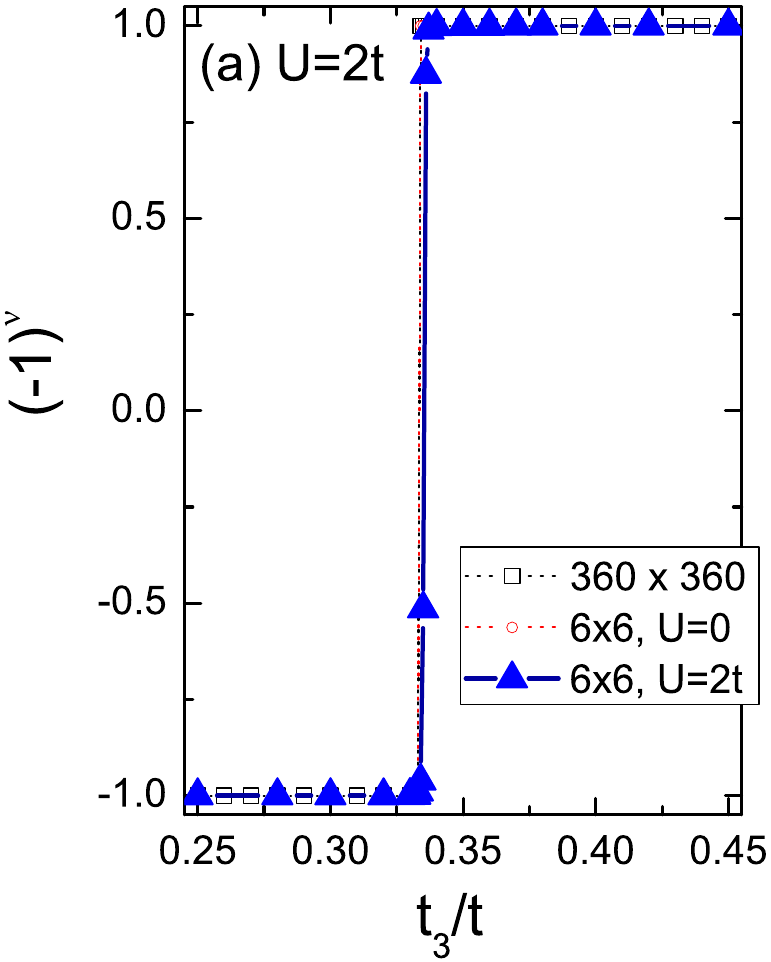}
      \end{minipage}
      \hfill
      \begin{minipage}[t]{0.3\textwidth}\centering
         \includegraphics[width=\textwidth]{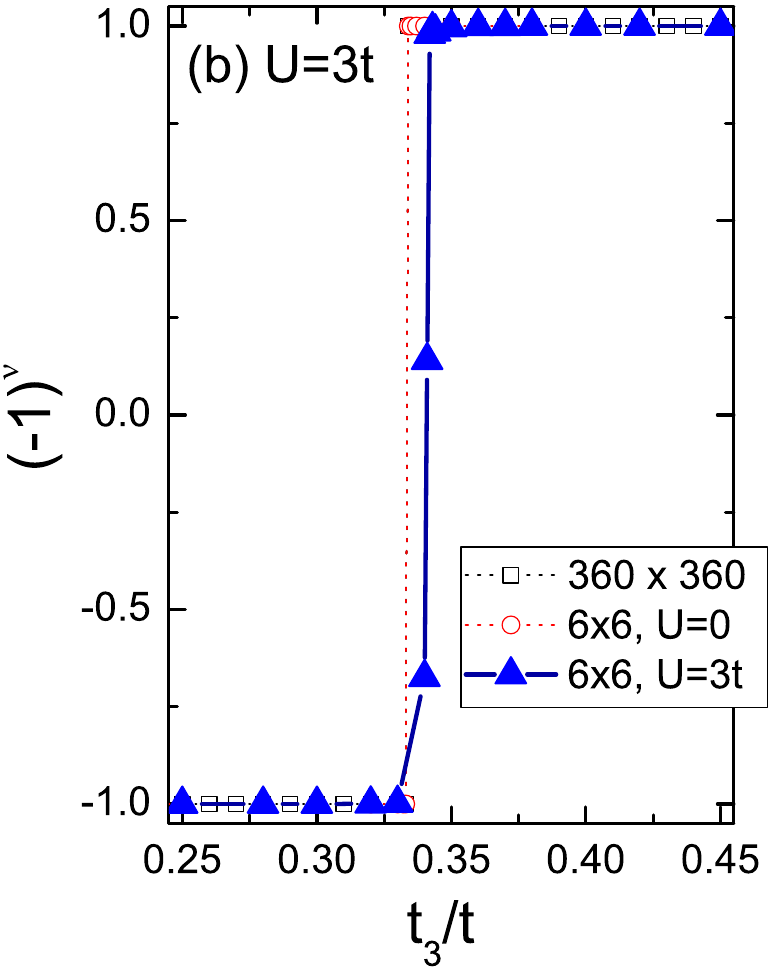}
      \end{minipage}
      \hfill
      \begin{minipage}[t]{0.3\textwidth}\centering
         \includegraphics[width=\textwidth]{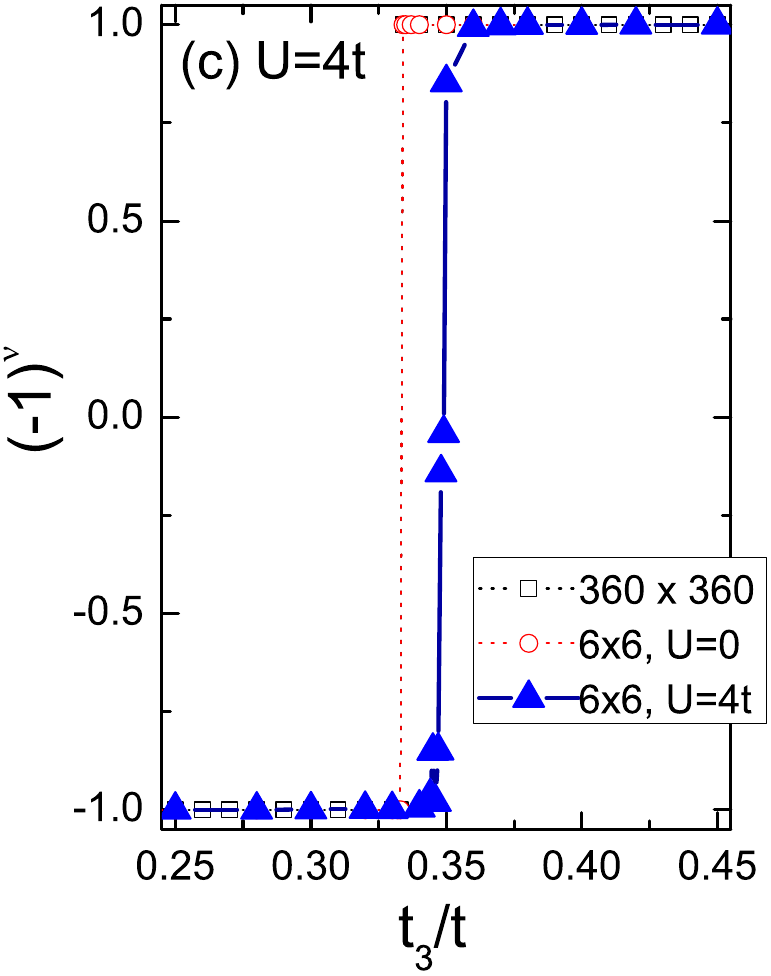}
      \end{minipage}\\
      \begin{minipage}[t]{0.3\textwidth}\centering
         \includegraphics[width=\textwidth]{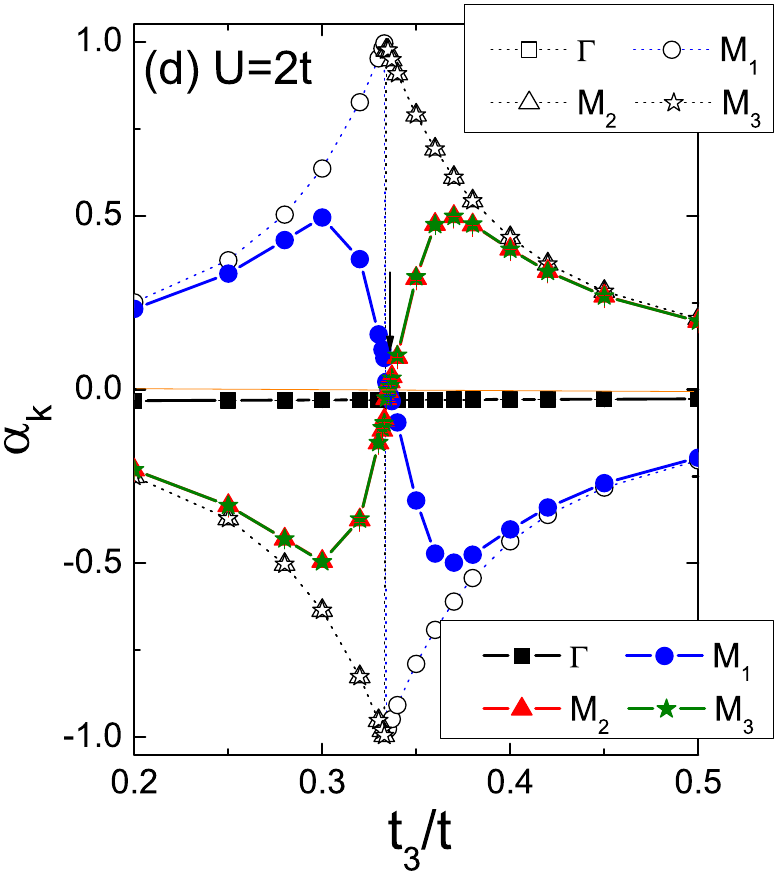}
      \end{minipage}
      \hfill
      \begin{minipage}[t]{0.3\textwidth}\centering
         \includegraphics[width=\textwidth]{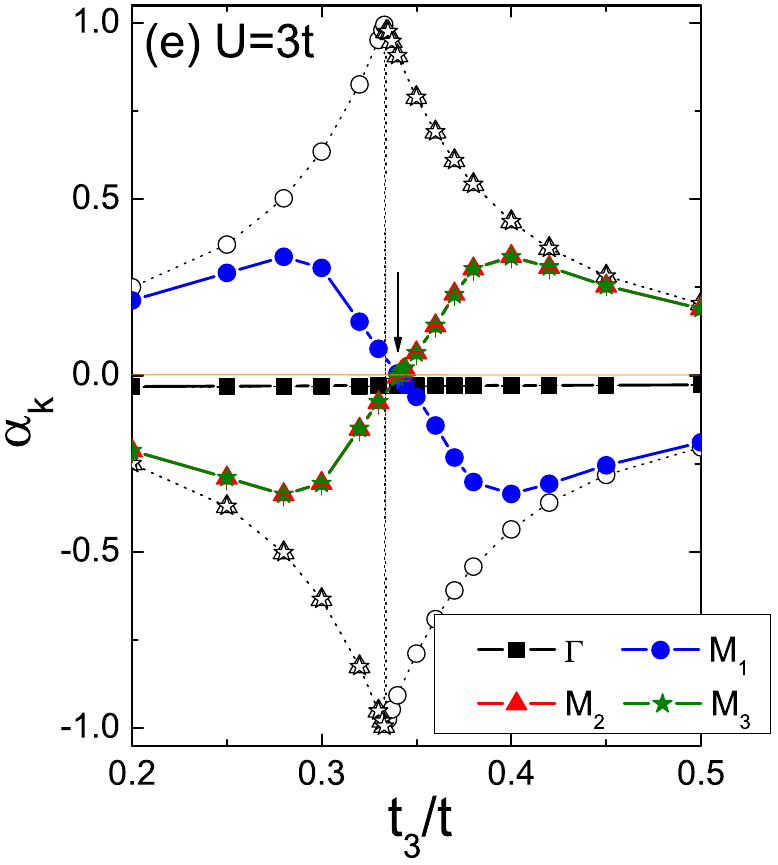}
      \end{minipage}
      \hfill
      \begin{minipage}[t]{0.3\textwidth}\centering
         \includegraphics[width=\textwidth]{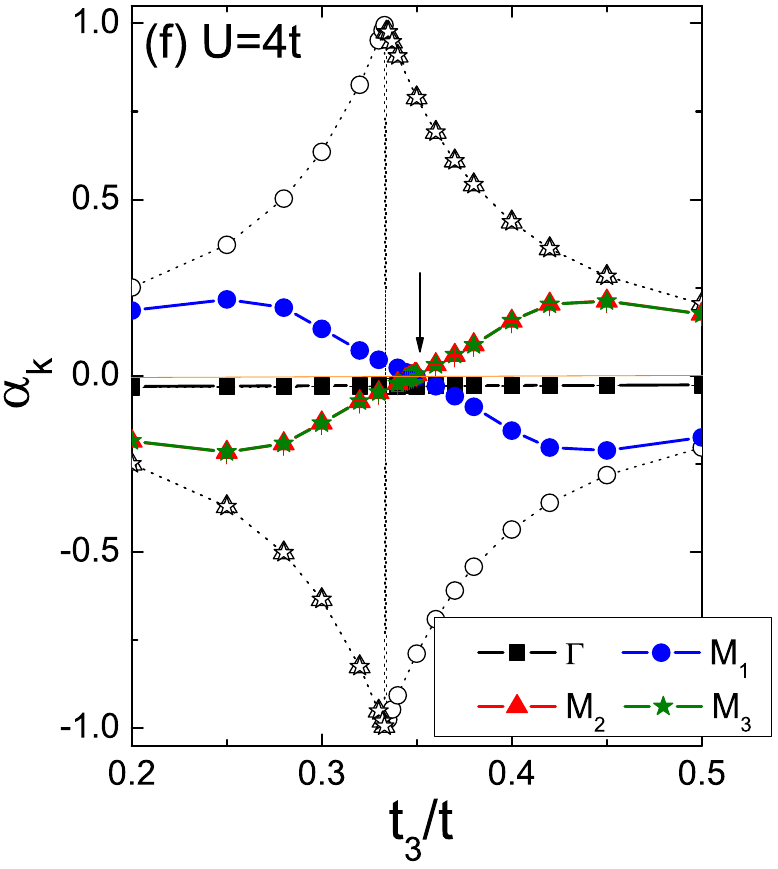}
      \end{minipage}
   \caption{(a)$-$(c) The $Z_2$ invariant at $U/t=2-4$ versus $t_{3}/t$ for
$\lambda=0.4t$. The black squares and red circles show the
noninteracting $Z_2$ invariant given by the tight-binding
calculations on a $L=200$ and by QMC simulations on a $L=6$ cluster.
The blue solid triangles depict the $Z_2$ invariant of the
generalized KMH model at $U \ne 0$. (d)$-$(f) show the proportional
coefficient $\alpha_{\mathbf k}$ determined from
Eq.~(\ref{eq:gamma}) from QMC simulations versus $t_{3}/t$. In these
plots, we use a convention that all the open symbols indicate
noninteracting cases whereas the solid symbols denote interacting
cases. The background orange line denotes zero. Errorbars have been omitted for clarity, or are smaller than the symbol size.
Adapted and reproduced with permission from Ref.~\onlinecite{Hung2013}.
Copyright 2013 American Physical Society.} \label{fig:Z2_coeff}
   \end{center}
\end{figure}

For the noninteracting case with $\lambda>0$, the critical value
$t^c_{3}=\frac{1}{3}t$  separates topologically non-trivial and
trivial phases. In Figs.~\ref{fig:Z2_coeff}, we present the $Z_2$
invariant as a function of $t_{3}$ for different values of $U$ for
$\lambda=0.4t$ on a $L=6$ cluster. Open and solid symbols denote the
noninteracting and interacting cases, respectively. For comparison,
we also show the noninteracting ($U=0$) $Z_2$ invariant from QMC
simulations on a $L=6$ cluster (open red circles) and from
tight-binding calculations on a $L=360$ cluster (open black
squares). Both results show that the $Z_2$ invariant varies at
$t^c_3=\frac{1}{3}t$, and confirm the accuracy of our small-size
QMC calculations in the noninteracting limit. In the topological
insulator phase ($t_3 < t^c_3$), only the $M_1$ point is parity odd
($\tilde{\eta}_{M_{1}}=-1$); the other three TRIM are parity even
($\tilde{\eta}_{\Gamma}=\tilde{\eta}_{M_{2,3}}=+1$), so
$(-1)^{\nu}=-1$ and $\nu=1$. Across the transition upon increasing
$t_{3}$, $\tilde{\eta}_{M_{1,2,3}}$ change parity. $\Gamma$ and
$M_1$ are parity even, whereas, $M_{2,3}$ are parity-odd, so
$(-1)^{\nu}=1$ and $\nu=0$.

Next let us turn to observe the interacting case ($U \ne 0$). In
order to avoid  invalidating Eq.~(\ref{eq:z2invariant}), we properly
choose the value of $U$ without inducing the magnetic
ordering. For $\lambda=0.4t$, we numerically confirm that in the
regime of $0.2t \le t_3 \le 0.45t$, $U=4t$ is still below the
critical interaction.\cite{Hung2013} The blue solid triangles in
Figs.~\ref{fig:Z2_coeff}(a)$-$(c) depict the dependence of the $Z_2$
invariant on $t_{3}/t$. In the presence of electronic correlations,
the parity properties of the TRIM still remain, and
Eq.~(\ref{eq:z2invariant}), to evaluate the $Z_2$ invariant, remains
well defined. This is because at moderate interaction the system
adiabatically connects to the QSH state, as long as the band gap
remains open. Note that here $\langle \tilde{\eta}_{\mathbf{k}}
\rangle=\pm 1$ is determined over thousands of QMC configurations,
with small statistical errors. At weak interaction, $U=2t$, the
phase boundary is estimated at $t_{3}=0.335t$, which only slightly
deviates from the noninteracting $t^{c}_3=\frac{1}{3}t$. By
increasing $U$, however, one can explicitly see that the critical
points start to move towards larger values, indicating that the
topological insulator phase is stabilized by the interactions. At
$U=3t$ and $4t$, the critical points are estimated at
$t^c_{3}=0.341t$ and $0.348t$, respectively. This indicates that
a slight shift ($\sim 10\%$) of the topological phase boundary
is driven by the Hubbard interaction. This is in contrary to the
correlation effects in the dimerized KMH model which we shall
discuss later.

It is interacting to note that, such a boundary shift originates
from quantum fluctuations, since no shift as a function of $U$
is observed in a static Hartree-Fock mean-field
approximation. As long as $U<U_c$ without inducing the
antiferromagnetic ordering, the Hartree-Fock result is the same as
the tight-binding calculation. Therefore, the QMC results can
efficiently capture the quantum fluctuations originating in the
interactions and one can study the topological phase under
electronic correlation accurately.

Next, we investigate how the single-particle Green's function
behaves during the topological phase transition. In the inversion
symmetric generalized KMH model, at the TRIM, the zero-frequency
Green's functions for each spin can be simply expressed in terms of
$\gbf{\sigma}^x$ as $G_{\sigma}({\mathbf{k}},0)=\alpha_{\mathbf{k}}
\gbf{\sigma}^x$ [cf. Eq.~(\ref{eq:gamma})]. In
Figs.~\ref{fig:Z2_coeff}(d)$-$(f), we show the proportionality
coefficient $\alpha_{\mathbf{k}}$ as a function of $t_{3}$ for
finite $U$. For comparison, $\alpha_{\mathbf{k}}$ in the
noninteracting case is also depicted. At $U=0$, we find the
universal relations, $\alpha_{M_2}=\alpha_{M_3}$ and
$\alpha_{M_1}=-\alpha_{M_2}$, for all values of $\lambda$ and
$t_{3}$. The values of $\alpha_{\Gamma}$ (denoted by black hollow
squares, covered by the solid squares) behave smoothly as $t_{3}$
passes through $t^c_3$. However, $\alpha_{{\mathbf{k}}}$'s of the
other TRIM are divergent at $t_{3}=\frac{1}{3}t$ and change signs at
the topological phase transition. This can be realized that at a
critical point, the gap closes at the TRIM, so the zero-frequency
Green's functions behave divergently on the poles and then change
signs.\cite{Gurarie2011} For any $\lambda$ and at $U=0$, the
location of the sign change is always at $t^{c}_3$, implying that
the behavior of $\alpha_{\mathbf{k}}$ can be another indication to
determine the topological phase transitions, like the $Z_2$
invariant.

Similarly, turning on the Hubbard interaction $U$, one can still
observe the sign change in $\alpha_{\mathbf{k}}$ at the topological
phase transitions. For finite $U$, within QMC simulation errorbars the
zero-frequency Green's functions retain their $\gbf{\sigma}^x$-like
form, and the universal relations $\alpha_{M_2} \simeq \alpha_{M_3}$
and $\alpha_{M_1} \simeq -\alpha_{M_2}$ still hold, independent of
the value of $U/t$. However, the positions where
$\alpha_{\mathbf{k}}$'s change signs move away from $\frac{1}{3}t$.
In Figs. \ref{fig:Z2_coeff}(d)$-$(f), the arrows label the location
of the sign changes in $\alpha_{\mathbf{k}}$, indicating the
topological phase boundaries in the interacting case. Clearly,
compared with the upper panels, Figs. \ref{fig:Z2_coeff}(a)$-$(c),
the locations for the sign change are consistent with the places
where the $Z_2$ invariant jumps.

\begin{figure}[tp]
   \begin{center}
      \begin{minipage}[t]{0.48\textwidth}\centering
         \includegraphics[height=5.8cm]{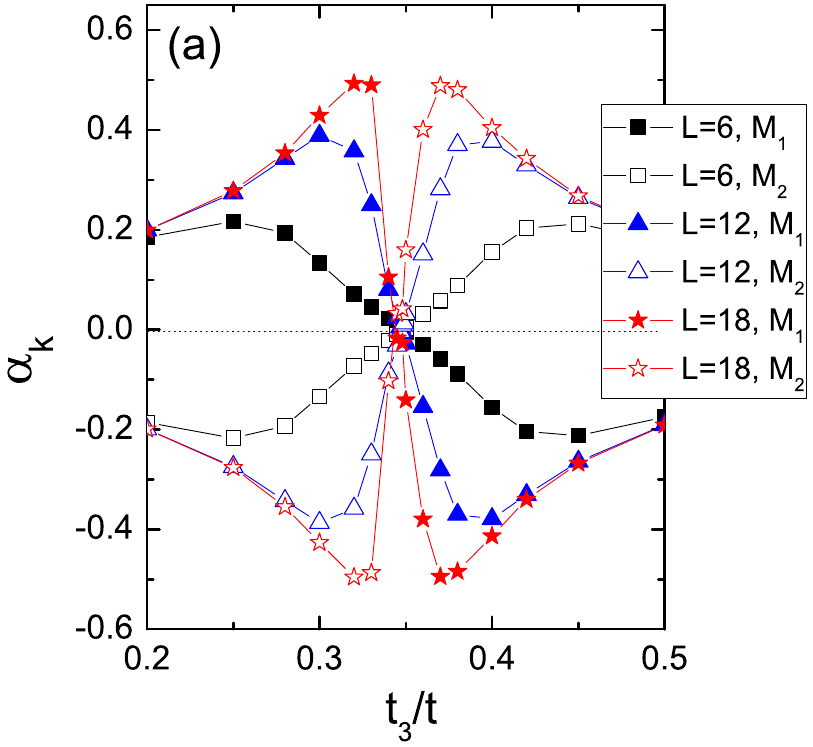}
      \end{minipage}
      \hfill
      \begin{minipage}[t]{0.48\textwidth}\centering
         \includegraphics[height=5.9cm]{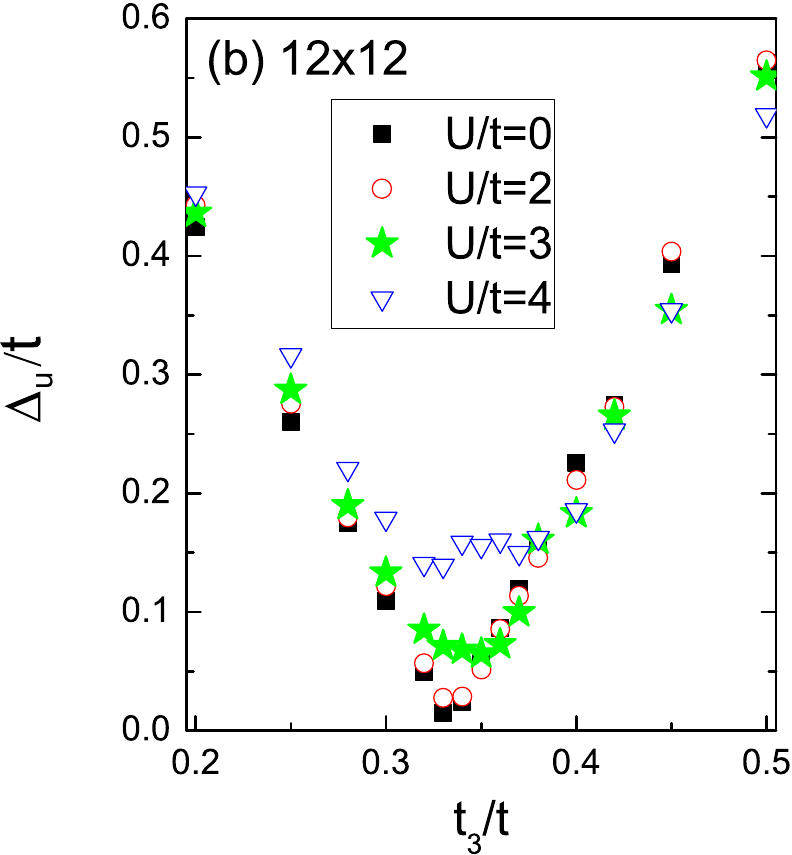}
      \end{minipage}
   \caption{(a) The
proportionality coefficient $\alpha_{\mathbf{k}}$ for
${\mathbf{k}}=M_{1,2}$ versus $t_{3}/t$ for $\lambda=0.4t$ on $L=6$, $L=12$ 
and $L=18$ clusters at $U=4t$. (b) Uniform single-particle gap 
$\Delta_\text{u}$ for different values of interaction: ${U/t=0,2,3}$
and $4$ with $\lambda=0.4t$ on the $L=12$ cluster. 
Errorbars have been omitted for clarity. Adapted and 
reproduced with permission from Ref.~\onlinecite{Hung2013}. 
Copyright 2013 American Physical Society.} \label{fig:coeffsize}
   \end{center}
\end{figure}

In Figs.~\ref{fig:Z2_coeff}(d)$-$(f), one can also observe how the
$\alpha_{\mathbf{k}}$ coefficients evolve upon increasing
interactions.  At larger $U$, the magnitude of $\alpha_{\mathbf{k}}$
is more suppressed. Although the sign change is still evident, the
sign-flip behavior becomes more smooth with stronger interaction.
This corresponds to a smeared phase boundary indicated by the $Z_2$
invariant changes in Figs. \ref{fig:Z2_coeff}(a)$-$(c). Such an
effect, however, will become less important upon increasing sizes.
Figure~\ref{fig:coeffsize}(a) shows the size dependence of the
coefficients $\alpha_{\mathbf{k}}$ at ${\mathbf{k}}=M_{1,2}$ versus
$t_3/t$. The spin-orbital coupling and interaction are fixed at
$\lambda=0.4t$ and $U=4t$. Note that $M_1$ has opposite parity to
$M_2$, so the coefficients have opposite sign at these momenta. Upon
the topological phase transition, both $\alpha_{\mathbf{k}}$ flip
signs. However, one can see that upon increasing sizes, the behavior
of $\alpha_{\mathbf{k}}$ near the $t^c_3$ is getting divergent,
which is observed in the noninteracting limit.

In addition to the values of $\alpha_{\mathbf{k}}$, one can also
observe a weak finite-size dependence on the locations of sign flip.
This implies that considering small clusters is able to identify the
topological phase transition. Interestingly, away from the
topological phase transitions, e.g., $t_{3}=0.2t$ and $0.5t$, the
coefficients $\alpha_{\mathbf{k}}$ for $U \ne 0$ seem to be
consistent with the noninteracting values. Therefore, interaction
effects in $\alpha_{\mathbf{k}}$ are most apparent as $t_{3}$
approaches the topological phase transition.

Besides the $Z_2$ invariant and associated the Green's function
behavior, monitoring the closing of the single-particle gap is also
another possible indicator for the topological phase transition.
However, the single-particle gap is subject to a stronger
finite-size effect. Figure~\ref{fig:coeffsize}(b) shows for a fixed system 
size ${L=12}$, that at ${U=0}$ and ${U=2t}$, one can clearly see the 
vanishing gap location. At ${U=3t}$ the behavior becomes less obvious but 
the minimum gap location is still visible. Furthermore at ${U=4t}$, the plot
cannot provide clear information to identify the location of the
topological phase transition. While this observation could be
systematically improved by performing finite-size analysis,
it is clear that the behavior of the single-particle gap is not as
sensitive as the topological invariant, since large lattices are
needed in order to perform reliable finite-size scaling.

\vspace{2em}\noindent\textit{Dimerized Kane-Mele-Hubbard model\vspace{1em}}\

\begin{figure}[tp]
    \centering
\includegraphics[width=\textwidth]{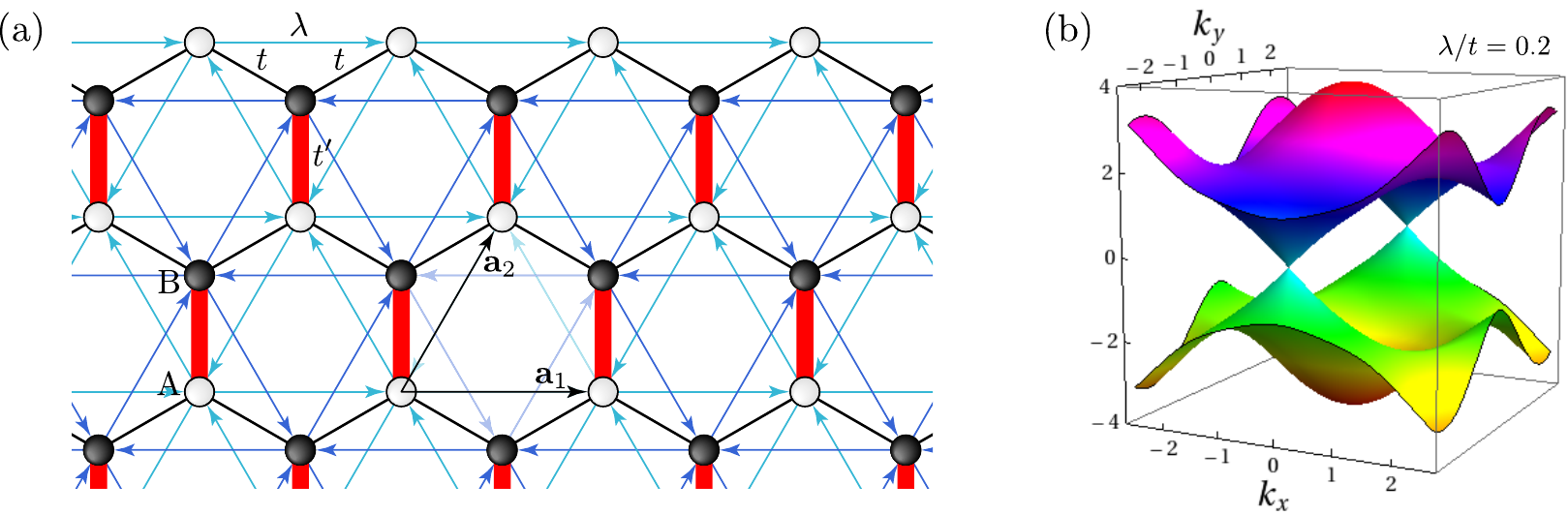}
   \caption{(a) The lattice geometry of the DKM model with the dimerized 
bonds $t'$ indicated by bold red lines and (b) the band structure of the 
DKM model at $t_c'=2t$. At the transition from the topological insulator 
${(t'<t_c')}$ to the band insulator ${(t'>t_c')}$ the single-particle 
gap closes at the TRIM $M_3$.
   \label{fig:DKMmodel}}
\end{figure}

\noindent Complementary to the the above mentioned GKM model, it has
been shown that the KM model with anisotropic nearest-neighbor
hopping can also exhibit a topological phase transition into a
trivial band insulator.\cite{Lang2013} This is the so-called
dimerized Kane-Mele (DKM) model and its Hamiltonian is given by
\begin{equation}
    H_\text{DKM} = - t\sum_{\langle i,j \rangle,\sigma}c_{i\sigma}^{\dagger}c_{j\sigma}
    - t'\sum_{\langle i,j \rangle,\sigma}c_{i\sigma}^{\dagger}c_{j\sigma}
    + \I\,\lambda\sum_{\langle\!\langle i,j \rangle\!\rangle, \alpha\beta} \nu_{ij}\,c^{\dagger}_{i\alpha}\sigma^{z}_{\alpha\beta}c_{j\beta} \;.
   \label{eq:DKMHam}
\end{equation}
Different from the GKM model, the DKM Hamiltonian only contains
nearest-neighbor hopping. However, one of the three nearest-neighbor
hoppings is chosen with a different amplitude $t'$ along the
direction $\gbf{\delta}_{1}=(0,1)$ compared to the other two along
the directions, as shown in Fig.~\ref{fig:DKMmodel} (a).

The Hamiltonian in momentum space can be written as ${H_\text{DKM}=
\sum_{\mathbf{k}}\Phi^{\dagger}_{\mathbf{k}}H(\mathbf{k})\Phi_{\mathbf{k}}}$, with
\begin{equation}
    H(\mathbf{k}) = \left( \begin{array}{cccc}
   \gamma_{\mathbf{k}} & -g'_{\mathbf{k}} & 0 & 0 \\
   -g^{'\ast}_{\mathbf{k}} & -\gamma_{\mathbf{k}} & 0 & 0 \\
   0 & 0 & -\gamma_{\mathbf{k}} & -g'_{\mathbf{k}} \\
   0 & 0 & -g^{'\ast}_{\mathbf{k}} & \gamma_{\mathbf{k}} \end{array} \right)\;.
\label{eq:dkmhamiltonianmatrix}
\end{equation}
The off-diagonal term $g'_{\mathbf{k}}$ is different from the corresponding 
term $g_{\mathbf{k}}$ in the Kane-Mele Hamiltonian $H_\text{KM}$ in 
Eq.~(\ref{eq:kmhamiltonianmatrix}) due to the anisotropic nearest-neighbor 
hopping $t'$. It is given by $g'_{\mathbf{k}}=t'\exp(\I\mathbf{k}\cdot\gbf{\delta}_{1}) 
+ t\,[\exp(\I\mathbf{k}\cdot\gbf{\delta}_{2})+\exp(\I\mathbf{k}\cdot\gbf{\delta}_{3})]$. 
The DKM Hamiltonian is also time-reversal symmetric, since 
Eq.~(\ref{eq:dkmhamiltonianmatrix}) it block diagonal in spin space, and 
$\gamma_{-\mathbf{k}}=-\gamma_{\mathbf{k}}$ and $g'_{-\mathbf{k}}=
g^{'\ast}_{\mathbf{k}}$. The dispersion of the DKM system follows from the 
eigenvalues of $H(\mathbf{k})$ and is given by $\varepsilon_\text{DKM}(\mathbf{k})=
\pm\sqrt{|g'_{\mathbf{k}}|^{2}+\gamma^{2}_{\mathbf{k}}}$.

\begin{figure}[tp]
\centering
\includegraphics[width=0.9\textwidth]{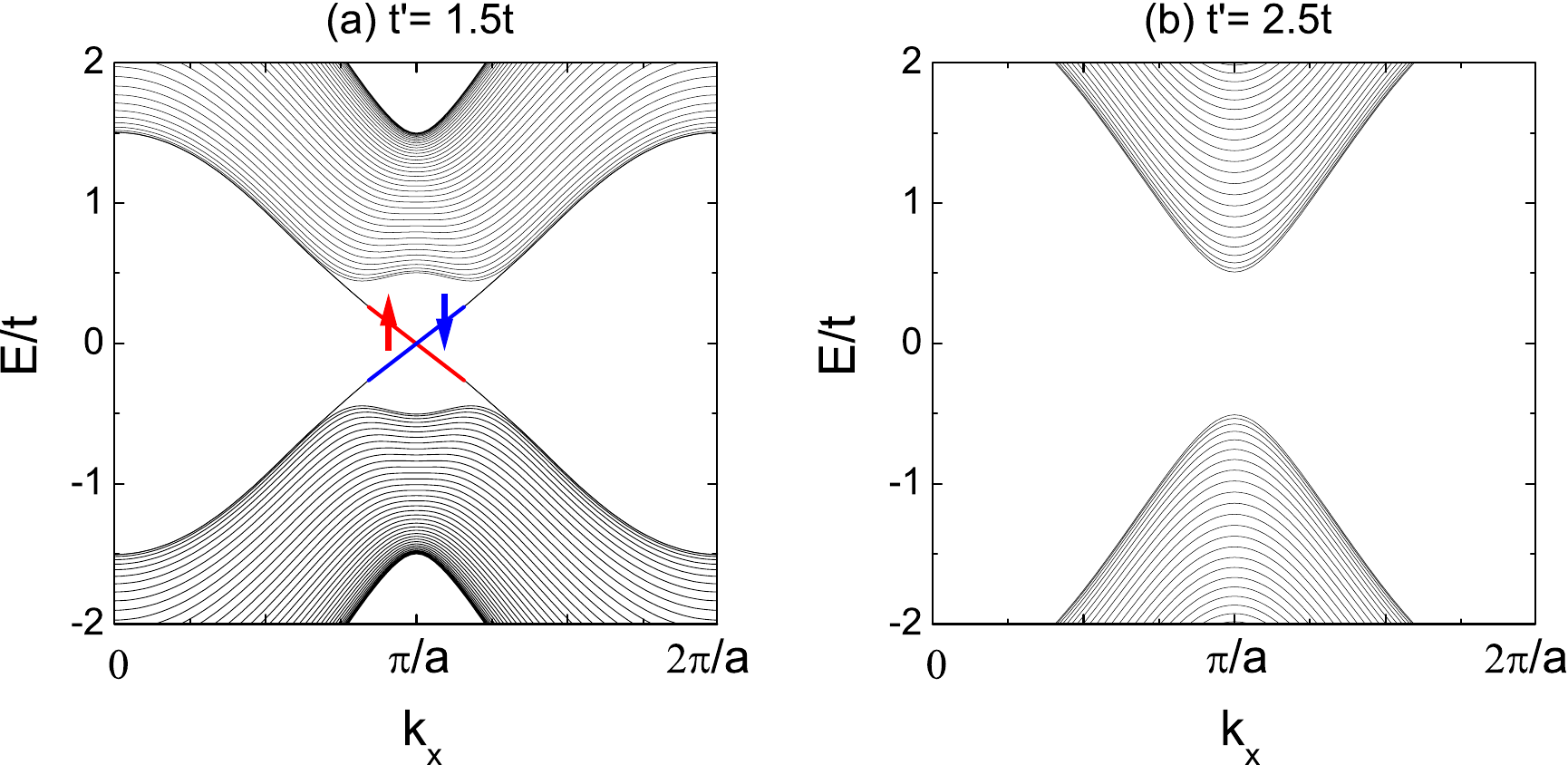}
   \caption{(Color online) The edge spectra for the noninteracting DKM
model at (a) $t'=1.5t$  and (b) $t'=2.5t$, for a $Z_2$ topological
insulator and a trivial insulator, respectively. $\lambda=0.2t$ is
used. Here, the anisotropic hopping $t'$ was introduced along the $\gbf{\delta}_2$
direction.} \label{fig:edgestate_dimer}
\end{figure}

As discussed in Sec.~\ref{sect:Z2formalism}, the $Z_2$ topological
invariant is identified by the zero-frequency Green function at the
four TRIM $G_{\sigma}(\gbf{\kappa},0)$, and is calculated according
to Eq.~(\ref{eq:z2invariant}) from any of the two spin sectors,
which together form a Kramer's pair at each TRIM. In the
noninteracting case, $U=0$, we obtain for finite values of
spin-orbit coupling $\lambda$, a change in the $Z_2$ invariant from
$\nu=1$ for $t' < 2t$ to $\nu=0$ for $t' > 2t$. This indicates the
topological phase transition from a $Z_2$ topological insulator (the
QSH insulator) to a trivial band insulator exactly at $t_c'=2t$. As
shown in Fig.~\ref{fig:DKMmodel}(b), at $t'=2t$, the single-particle
gap closes at the $M_3$ point [compare with
Fig.~\ref{fig:kmtopologicalphasetransition} and
Fig.~\ref{fig:gkmtopologicalphasetransition} (b)]. We present the
corresponding edge spectra for the DKM model in
Fig.~\ref{fig:edgestate_dimer}. Panel (a) shows a helical mode at
$k_x=\pi$ implying that it is a $Z_2$ topological insulator with
spin Chern number $C_{\sigma}=\pm 1$. The spectrum in panel (b)
exhibits no edge modes corresponding to $C_{\sigma}=0$ [compare with
Fig.~\ref{fig:GKMedgestate}(b)].

\begin{figure}[tp]
\centering
\includegraphics[width=\textwidth]{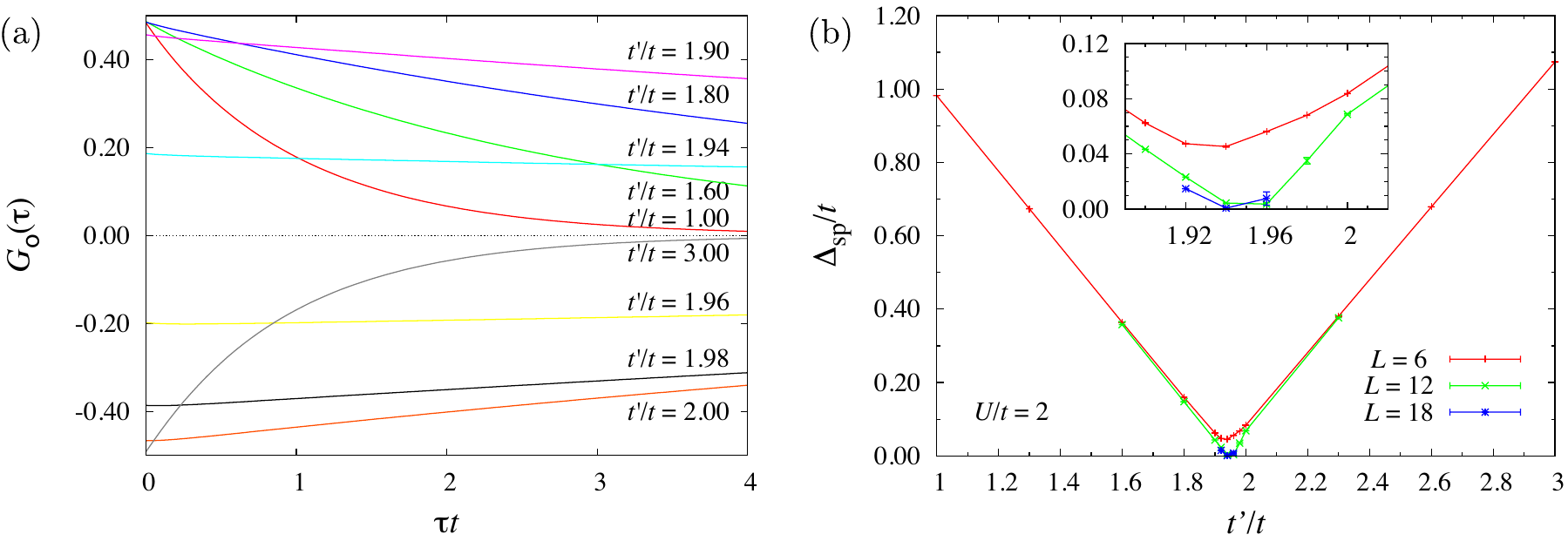}
\caption{(a)
Off-diagonal component of the Green function at $M_3$ for a system
of size $6\times6$, $U/t=2$ and $\lambda/t=0.2$ at various values of
$t'/t$. The area under the Green's function, which is proportional
to the $Z_2$ invariant changes from positive to negative between
$U/t=1.94$ and $U/t=1.96$. Errorbars are of the order of the
linewidth. (b) Evolution of the single-particle gap
$\Delta_\text{sp}$ at the $M_3$ point for different system sizes as
a function of $t'/t$ for $U/t=2$ and $\lambda/t=0.2$. In accordance
with transition indicated by the Green's function, the single
particle gap closes at the topological phase transition from the
topological insulator to the dimerized, trivial insulator. Adapted
and reproduced with permission from Ref.~\onlinecite{Lang2013}.
Copyright 2013 American Physical Society.} \label{fig:TCLFig3_4}
\end{figure}

Next let us consider the interacting case with the Hamiltonian
$H_\text{DKM}+H_U$, where $H_U  = \frac{U}{2} \sum_{i} (n_{i}-1)^2$,
and $n_{i}=\sum_{\sigma}n_{i,\sigma}$. In
Fig.~\ref{fig:TCLFig3_4}(a), the imaginary time dependence of the
off-diagonal component of the Green's function at the $M_3$ point
$G_{o}(\tau):=[G_{\uparrow}((M_{3},\tau)]_{AB}$ is shown. The area
under $G_{o}$ corresponds to the coefficients
$\alpha_{\gbf{\kappa}}$ [cf. Eqs.~(\ref{Gstructure2}) and
(\ref{eq:gamma})], hence a change in the $Z_2$ topological invariant
can be related to sign change of the area under $G_{o}(\tau)$. As
can be seen in Fig.~\ref{fig:TCLFig3_4}(a), for ${U/t=2}$ and
$\lambda/t=0.2$, this change occurs between ${t'/t=1.94}$ and
$1.96$, and correspondingly, the $Z_2$ invariant $\nu$ jumps from
$1$ to $0$. This means the topological-to-trivial band insulator
transition occurs at a value of ${t'/t=1.95(1)}$ -- smaller than in
the noninteracting case, where the critical values is ${t_c'=2t}$.
This can be understood as the consequence of the super-exchange
induced by the Coulomb repulsion $U$ which favors the singlet
formation on the $t'$-bonds. Similar to the GKM-Hubbard model, the
topological transition in DKM-Hubbard model is also associated with
the closing of the single-particle gap $\Delta_\text{sp}$ at $M_3$,
which may be obtained from the decay in imaginary time of the
diagonal Green function $[G_{\uparrow}(M_{3},\tau)]_{AA}\propto
\exp(-\tau\Delta_\text{sp}(M_{3}))$ and is shown in
Fig.~\ref{fig:TCLFig3_4}(b). The gap closes at ${t'/t=1.95(1)}$ and
thus supports the fact that the $Z_2$ invariant correctly captures
the interaction effects which lead to destabilize the topological
phase with respect to the noninteracting case.

\subsubsection{Limitations of the $Z_2$ Invariant in QMC simulations}
\label{sect:QMClimitation}

\begin{figure}[tp]
\centering
\includegraphics[width=\textwidth]{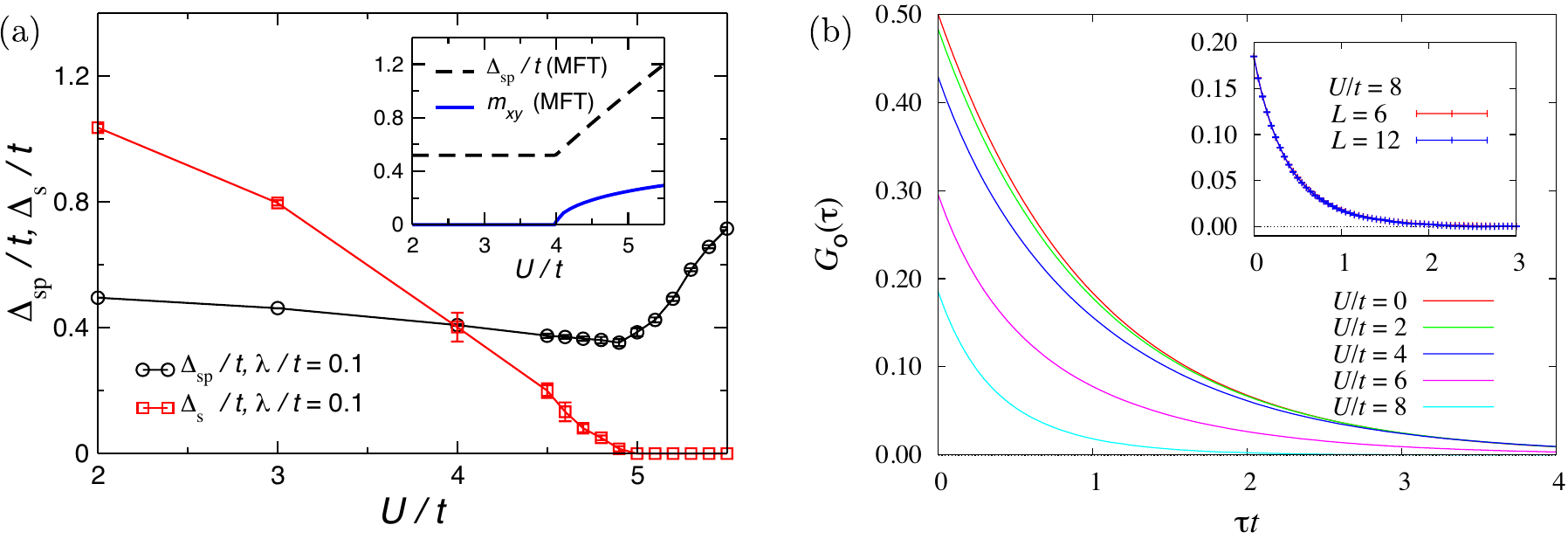}\caption{(a)
Single-particle gap $\Delta_\text{sp}$ and spin gap $\Delta_{s}$ as
a function of $U/t$ at $\lambda/t=0.1$ for the KMH model. The values
shown were obtained from an extrapolation of the finite size results
to the thermodynamic limit. The dip in $\Delta_\text{sp}$ and the
closing of $\Delta_{s}$ are consistent with the topological
insulator to antiferromagnetic Mott insulator transition at
$U/t\approx 5$. The inset shows the mean field results for the
single-particle gap and the magnetic order parameter. (b)
The off-diagonal component of the Green function in the KMH model 
at the $M_3$ point for $L=6, \lambda=0.2t$
and different values of $U$ from $U=0$ to $U=8t$ (top to bottom).
Across the transition from QSH insulator to the antiferromagnetic
Mott insulator the Green function remains qualitatively unchanged.
Errorbars are of the order of the line width and have been omitted
for clarity. Inset shows the Green functions have very little finite
size dependence. Adapted and reproduced with permission from
Ref.~\onlinecite{Hohenadler2012,Lang2013}. Copyright 2013 American
Physical Society.} \label{fig:ZYMgaps_TCLFig5}
\end{figure}

In the case, where the onsite Coulomb repulsion is large enough, previous
studies\cite{Lang2011,Zheng2011,Hohenadler2012} have shown that the
system enters a transverse antiferromagnetically ordered
Mott-insulating phase. With the onset of magnetic order, the
time-reversal symmetry is spontaneously broken. The phase transition
from topological insulator to the antiferromagnetically ordered
phase at a fixed value of $\lambda$ by increasing $U/t$, is however,
not accompanied by the closing of the single-particle gap. As shown
in Fig.~\ref{fig:ZYMgaps_TCLFig5}(a) for the KMH model at
$\lambda/t=0.1$ and as a function of $U/t$, the single-particle gap
merely exhibits a cusp at the transition point, ${U\approx 5t}$.
Since the antiferromagnetically ordered Mott-insulating phase breaks
the SU(2) spin rotational symmetry and Goldstone modes emerge in the
thermodynamic limit, it is actually the spin gap, defined from
dynamic spin-spin correlation function,
$S^{+-}_{AF}(\tau)=\frac{1}{L^{2}}\sum_{i,j}\langle
S^{+}_{i}(\tau)S^{-}_{j}(0)\rangle \propto \exp(-\Delta_s\tau)$,
that closes at the transition point, ${U\approx 5t}$. This
transition from the topological insulator to the antiferromagnetical
Mott insulator is induced by collective excitations at the
two-particle level. The $Z_2$ topological invariant based on the
zero-frequency Green can only capture physics in the single-particle
sector, and hence fails to detect this transition. This can also
be seen in Fig.~\ref{fig:ZYMgaps_TCLFig5}(b), for the KMH
model, where $G_{o}(M_3,\tau)$ is shown for different values fo
$U/t$ at $\lambda/t=0.2$.\cite{Lang2013} The magnetic
transition happens near $U\approx5t$, however the off-diagonal
component of the Green's function exhibits no qualitative change for
increasing interactions.

\begin{figure}[tp]
   \begin{center}
      \begin{minipage}[t]{0.49\textwidth}\centering
         \includegraphics[height=5.cm]{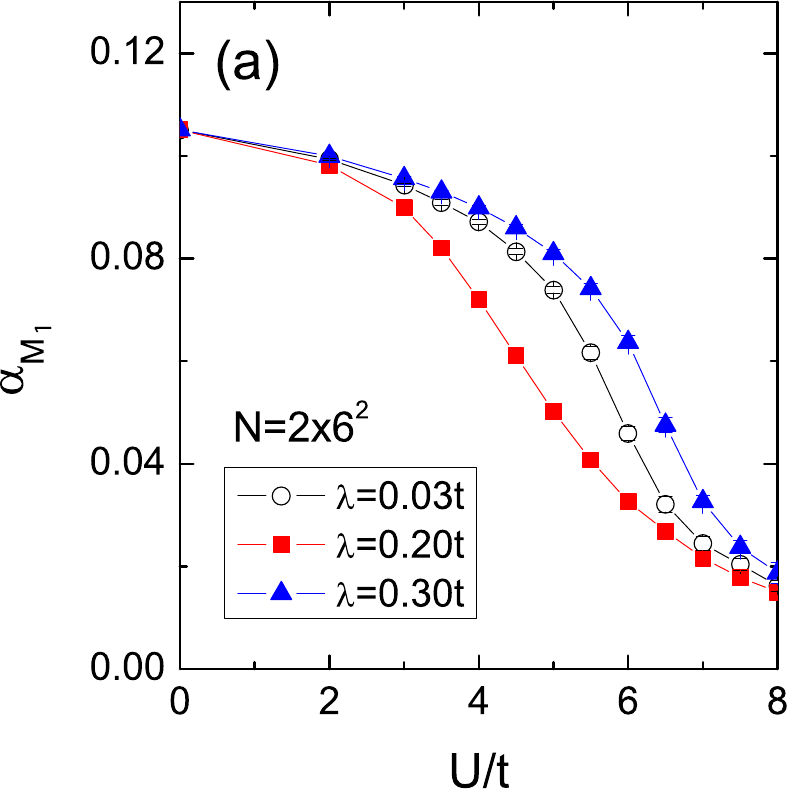}
      \end{minipage}
      \hfill
      \begin{minipage}[t]{0.49\textwidth}\centering
         \includegraphics[height=5.cm]{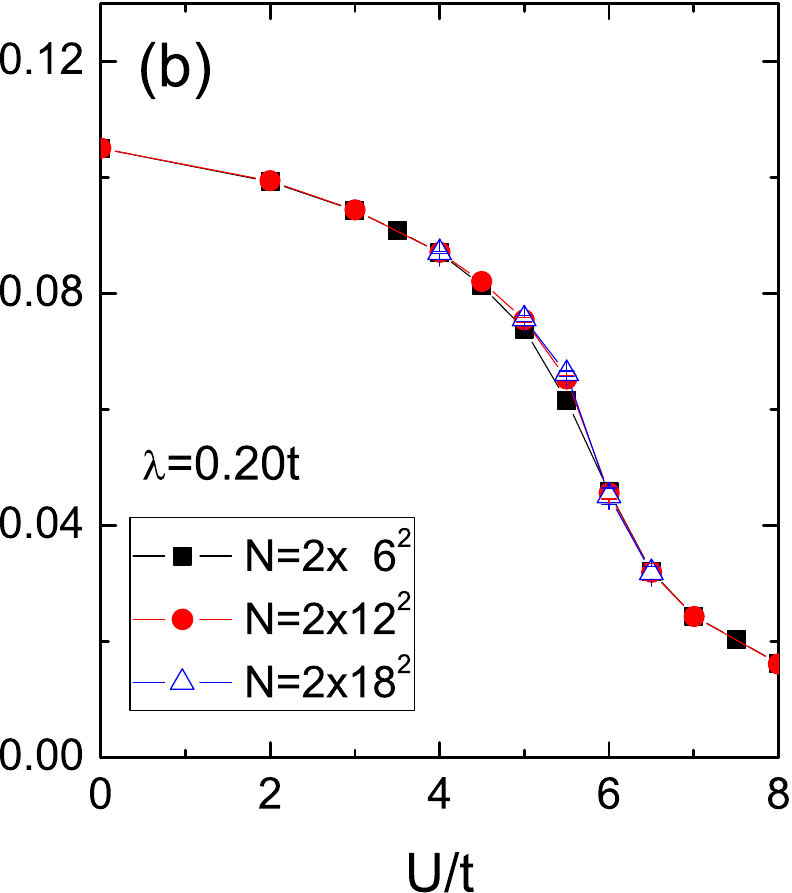}
      \end{minipage}
   \caption{The proportionality coefficient $\alpha_{\mathbf{k}}$ 
at $M_1$ versus $U/t$ in the KMH model. (a) The comparison
between $\lambda=0.03t$, $0.2t$ and $0.3t$ on a $L=6$ cluster. (b)
The size dependence comparison at $\lambda=0.2t$.} \label{fig:KMH_coeff}
   \end{center}
\end{figure}

In contrast to the sign change of  $G_{o}(M_3,\tau)$ as one scan
$t'/t$ acrossing the topological phase transition shown in
Fig.~\ref{fig:TCLFig3_4}(a),  $G_{0}(M_3,\tau)$ in
Fig.~\ref{fig:ZYMgaps_TCLFig5}(b) remains positive as the
interaction strength $U/t$ varies across the topological insulator
to antiferromagnetic Mott insualtor transition point. We emphasize
that this does not appear to be a mere finite size effect, as can be
seen in the inset of Fig.~\ref{fig:ZYMgaps_TCLFig5}(b), where we
compare the QMC data at $U/t=8$ for two different system sizes,
$L=6$ and $12$, and fall perfectly on top of each other. The results
are seen to indeed be converged in the finite sizes we have studied.
The associated proportionality coefficients $\alpha_{\gbf{\kappa}}$
which correspond to the area under $G_{o}$ are shown in Fig.~\ref
{fig:KMH_coeff} as a function of $U$ and verify the issue.
Figure~\ref{fig:KMH_coeff}(a) shows that, although the values of
$\alpha_{\gbf{\kappa}}$ decay near the expected $U_c$, there is no
sign change in $\alpha_{\mathbf{k}}$ upon tuning $U$ through the
critical value for all values of $\lambda$ we have studied.
Fig.~\ref{fig:KMH_coeff}(b) considers finite size dependence of
$\alpha_{\gbf{\kappa}}$ versus $U$ at $\lambda=0.2t$. Again, the
absence of strong finite size effects is obvious.

Both, Fig.~\ref{fig:ZYMgaps_TCLFig5} and Fig.~\ref{fig:KMH_coeff}
suggest that the $Z_2$ invariant $\nu$ stays constant
across the topological insulator to antiferromagnetic Mott insulator
phase transition. Once the system enters the antiferromagnetic
ordered phase, the time-reversal and inversion- (sublattice)
symmetries of the Hamiltonian are spontaneously broken. This happens
at the two-particle level and hence cannot be monitored by the
single-particle Green's function, on which the calculation of the
$Z_2$ invariant is based. Strictly speaking, the spontaneous breaking of spin- and
time-reversal-symmetry applies only to the thermodynamic limit. While
order parameters can acquire finite values on finite size lattices,
no symmetry, neither continuous nor discrete, can be spontaneously
broken unless in the thermodynamic limit. Hence the $Z_2$ invariant
formalism Eq.~(\ref{eq:z2invariant}) is  formally well defined on
finite-size lattices. While spontaneous symmetry breaking in
the ordered phase implies a degenerate ground state subspace in the
limit of infinite system size, in finite-size simulation the
ground state is given by the linear combination of equal weight 
of states from this manifold. In our case all the different spin-orientated 
symmetry breaking states have equal weight, such that 
measurements on this ground state are not able to distinguish any 
preferred magnetic ordering. 

Even if the interaction strength is large
enough to trigger spontaneous symmetry breaking in the
thermodynamic limit, our results in Figs.~\ref{fig:ZYMgaps_TCLFig5}
and \ref{fig:KMH_coeff} do not indicate any qualitative change with 
increasing system size. Thus, limited to the single-particle sector, 
the $Z_2$ topological invariant is still blind with respect to collective
excitations and the associated spontaneous symmetry breaking. Its
usage as a reliable indicator of the topological nature of phases
(or lack thereof) should therefore be restricted to cases where the
transition from the topological insulator to the topological trivial
phase is also indicated by the closing of the single-particle gap.
In this regard, it does not really matter whether the vanishing of
the single-particle excitation gap at the critical point is due to
the underlying physics, or the artifacts associated with the method
used. Indeed, the successful application of the $Z_2$ topological
invariant has been shown in, e.g., correlated electron systems in
one dimension using the numerically exact
time-dependent density matrix renormalization group (DMRG)
approach;\cite{Manmanna12} For two-dimensional interacting systems
in the approximative approaches using mean-field
theory\cite{Rachel2010} and the variational cluster approximation
(VCA)\cite{Budich12a} applied to the KMH model; Dynamical mean-field
theory (DMFT) has been employed to study the interaction-driven
transition between topological states in a Kondo
insulator\cite{Werner13} and cluster DMFT to study model for
three-dimensional correlated complex oxides, the pyrochlore
iridates.\cite{Go12} In these cases the $Z_2$ topological invariant
still allows for highly accurate determination of the
critical point. Recently, topological invariants
expressed in terms of ground state wavefunction are proposed for
topological insulators,\cite{wang2013wavefunction} which are valid
in the presence of arbitrary interaction. From the current numerical
perspective, the Green's function remains the most efficient
approach within the realm of its validity.

\subsection{The Spin Chern Number}
\label{sect:spinChern}

In addition to the $Z_2$ invariant, the topological insulators can
also be characterized by the spin Chern number (cf.
Sec.~\ref{sect:QSHI}), that is the Chern number for one spin flavor.
It has been shown that without the Rashba coupling, the
spin Chern number is still a good indicator for the QSH
state.\cite{Sheng2005,Sheng2006} On one hand, the Chern number can
be obtained from a many-body wave function with twisted boundary
conditions.\cite{niu1985} On the other hand, the spin Chern number
can be evaluated in terms of the projection operators
as\cite{Avron83}
\begin{eqnarray}
   C_{\sigma}
      & = & \frac{1}{2\pi\I}\int_{\text{B.Z.}} \text{Tr}\Big\lbrace P_{\sigma}({\mathbf k}) \Big[\partial_{k_x} P_{\sigma}({\mathbf k}) \partial_{k_y}P_{\sigma}({\mathbf k}) \nonumber \\
         &   & \quad\quad\quad\quad\quad\quad\quad\quad\, -\,\partial_{k_y} P_{\sigma}({\mathbf k}) \partial_{k_x}P_{\sigma}({\mathbf k}) \Big] \Big\rbrace\; dk_x dk_y\;,\label{eqn:chernnumb}
\end{eqnarray}
where $P_{\sigma}({\mathbf
k})=\sum_{u_n>0}|u_n(\mathbf{k})\rangle\langle u_n(\mathbf{k})|$ is
the spectral projector operator constructed using the Bloch
eigenvectors at ${\mathbf k}$  with energies below the Fermi energy
$\varepsilon_\text{F}$, i.e., $E_n({\bf k})<\varepsilon_\text{F}$.
Although the formalism was first proposed for noninteracting
systems, we can also compute the interacting spin Chern number with
the QMC method. In an analogy with the $Z_2$ invariant, the Bloch
eigenstates are replaced with the R-zero eigenvectors of the
zero-frequency Green's functions
$|\mu_{n}\rangle=|\mu_{n}(\mathbf{k},0)\rangle$, and then
\begin{equation}
   P_{\sigma}({\bf k})=\sum_{\mu_n>0}|\mu_n\rangle\langle\mu_n|\;,
\end{equation}
where choosing $\mu_i>0$ corresponds to selecting occupied bands
$E_n<\epsilon_F$, i.e. R-zero of the $G_{\sigma}({\bf k},0)$. In
finite-size systems, the integration over the Brillouin zone is
substituted by summation of discrete momentum points. For a $N=L
\times L$ lattice grid with spacing $h$, we can approximate the
$\partial_{k_x}P_{\sigma}(\mathbf{k})$ and $\partial_{k_y}P_{\sigma}(\mathbf{k})$ as
\begin{eqnarray}
   \partial_{k_x}P_{\sigma}(\mathbf{k}) &\approx& \frac{P_{\sigma,i+1,j}-P_{\sigma,i-1,j}}{2h} \;,\nonumber\\
   \partial_{k_y}P_{\sigma}(\mathbf{k}) &\approx& \frac{P_{\sigma,i,j+1}-P_{\sigma,i,j-1}}{2h} \;.
\end{eqnarray}
Thus, the spin Chern number in Eq.~(\ref{eqn:chernnumb}) can be
approximated as\cite{Hung2013b}
\begin{eqnarray}
   C_{\sigma}&\approx&\frac{1}{2\pi\I}\sum_{i,j=1}^{N}\frac{P_{\sigma,i,j}}{4}\big([P_{\sigma,i+1,j},P_{\sigma,i,j+1}]+[P_{\sigma,i,j+1},P_{\sigma,i-1,j}] \nonumber\\
   & & \quad\quad\quad\quad\quad\quad +\;[P_{\sigma,i-1,j},P_{\sigma,i,j-1}]+[P_{\sigma,i,j-1},P_{\sigma,i+1,j}]\big) \;.
\end{eqnarray}
Under such a construction, the evaluation of the spin Chern number
in the QMC simulations might look like subject to strong finite-size
effect and an quantized value of it is not guaranteed. However, 
we will demonstrate that, although subject to finite 
size effects, a jump in $C_{\sigma}$ can be clearly 
observed across the topological phase transition. This suggests that the
spin Chern number is a reliable means to detect topological properties 
even in the interacting systems.

\begin{figure}[tp]
   \begin{center}
      \begin{minipage}[t]{0.62\textwidth}\centering
         \includegraphics[width=\textwidth]{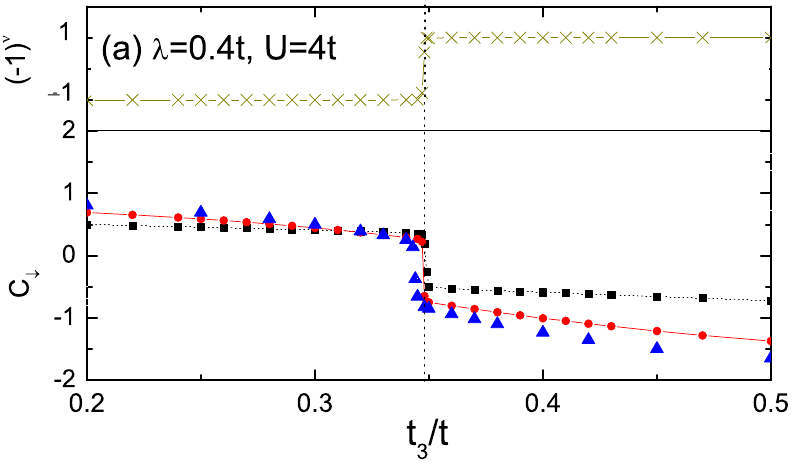}
      \end{minipage}
      \hfill
      \begin{minipage}[t]{0.37\textwidth}\centering
         \includegraphics[width=\textwidth]{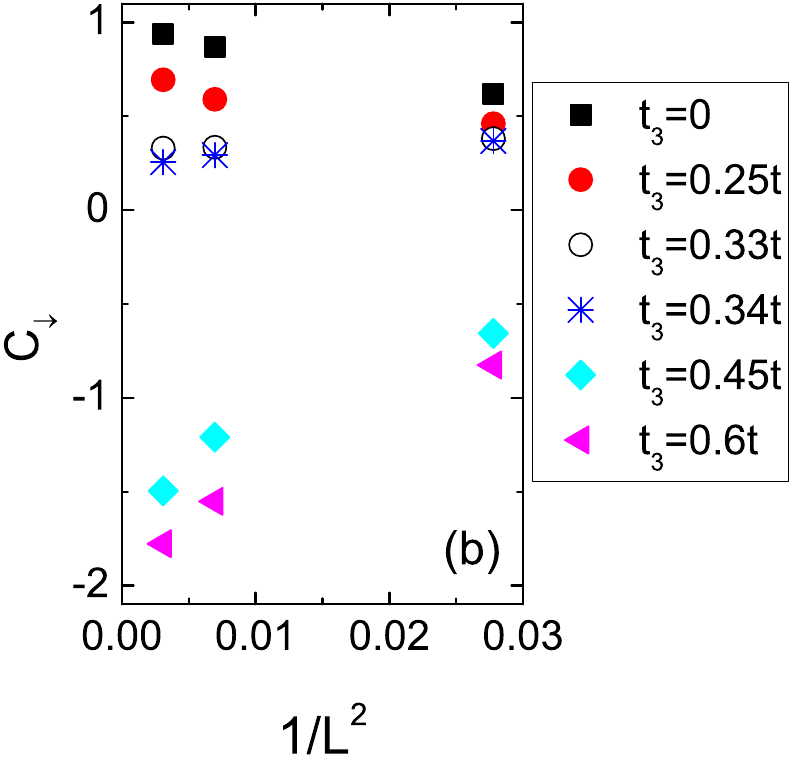}
      \end{minipage}
   \caption{(a) The $Z_2$ invariant $(-1)^{\nu}$ (upper
panel, for $L=12$ only) and spin Chern number $C_{\downarrow}$
(lower panel) for the GKM-Hubbard model as a function of $t_{3N}/t$
at $U=4t$. The spin-orbit coupling $\lambda=0.4t$ and the systems
sizes are chosen as $L = 6$ (squares), $12$ (circles) and $18$
(triangles). (b) The tentative finite-size scaling of
$C_{\downarrow}$ at $U=4t$. Errorbars have been omitted for clarity.} \label{fig:GKMspinchern}
    \end{center}
\end{figure}

As an example we compute the spin Chern number in the GKM model with 
interactions.\cite{Hung2013b} Figure \ref{fig:GKMspinchern}(a) shows the 
comparison of the $Z_2$ invariant $(-1)^{\nu}$ and the spin Chern number 
in the GKM-Hubbard model at $U=4t$ 
as a function of $t_{3}/t$. Although the spin Chern number is poorly quantized, 
in particular as $t_{3}$ is approaching to the transition (where
increasingly more Monte Carlo samples are needed to recover the
time-reversal symmetric relation $C_{\uparrow}=-C_{\downarrow}$
within statistical errors), a jump in $C_{\sigma}$ can be clearly
seen. 

This suggests that the spin Chern number is also reliable
to detect the topological phase transition at the interacting level.
Moreover, the poor quantization in the spin Chern number is
attributed to strong finite-size effects. The expected quantized
value can still be identified upon approaching to larger system
sizes. As an example, a tentative finite-size analysis
of the spin Chern number is shown in Fig.~\ref{fig:GKMspinchern}(b). 
Away from the topological phase transition, the quantized spin Chern 
number behavior can still be captured in the thermodynamic limit. 
One can see that in the small $t_3$ regime, $C_{\downarrow}=1$ 
whereas in the large $t_3$ limit, $C_{\downarrow}=-2$. Thus we can 
still distinguish the discrepancy during the topological phase 
transition at the interacting level. For details we refer the reader 
to Ref.~\onlinecite{Hung2013b}. Note, that based on the single-particle 
Green's function, the use of the spin Chern numbers to characterize 
the topological nature of a phase underlies the same limitations as 
the $Z_2$ invariant discussed in Sec.~\ref{sect:QMClimitation}.
The major advantage over the $Z_2$ topological invariant, evaluated at the 
TRIM, is that the spin Chern number can be even applied in systems 
\textit{without} inversion symmetry.

\section{Conclusion and Outlook}

The main objective of this review is to show how to detect the $Z_2$
topological nature of the KM model in the presence of electron correlations 
within QMC simulations. To this end, we discuss three
specific approaches and their application within the determinant quantum Monte
Carlo technique, with which the KMH model can be studied in
a unbiased manner. One special merit of the half-filled
KMH model is that the time-reversal symmetry and the
particle-hole symmetry allow for simulations free of the minus-sign problem --
hence the interplay of the topological properties of
the system with electronic interactions can be studied numerically exact.

The first approach is the idea of characterizing the
topological phase transition with magnetic flux insertion. We showed
that the $Z_2$ invariant can quantify the fluxon induction ($Z_2$
pumping) due to pairs of fluxes threaded through plaquettes of the system.

Next we explained in detail the idea of evaluating the $Z_2$ invariant in
terms of zero frequency Green's function. Due to the inversion
symmetry in the KMH systems, one only needs to evaluate the parity of 
the eigenstates of the zero-frequency Green's function at the time-reversal
invariant momenta. We provide two examples models, the generalized KMH
model and dimerized KMH model, and show that in the interacting case
the formulation of the $Z_2$ index evaluation can be easily calculated
from zero-frequency Green's functions. Both models
indicate a topological phase transition upon varying tight-binding
parameters. We show that the correlation effects stabilize the $Z_2$
order in the generalized KMH model, but destabilize it in the
dimerized KMH model. Although the $Z_2$ invariant could be used to 
successfully within QMC simulations for the KMH model systems, it is subject
to limitations. We discuss the quantum phase transition from topological insulator
to the magnetic insulator, which occurs at the two-particle level,
i.e., susceptibilities diverge, and where the zero frequency single-particle
Green's function is not able to capture such transitions. New ideas and 
formalisms are needed in these situations.

The third approach, accessible to detect topological phase transitions within
QMC simulations, is to directly measure the spin Chern number. Although under
strong finite-size effects, the spin Chern number measurement proves to present
another useful topological quantity in correlated topological insulators and 
can be even applied in systems intrinsically without inversion symmetry.

The three approaches above have their individual benefits and limitations, but
allow us to gather unbiased information on the topological nature in the correlated
quantum spin Hall system from simulations. Besides the KMH model, topological
phase transitions can be realized in more general models and systems, which do not 
retain particle-hole, inversion-, or even the time-reversal symmetry. In realistic
condensed matter materials, which might host topological states (topological insulator,
axion insulator, topological Mott insulator, or topological supercoductor),\cite{Witczak-Krempa13} 
electron correlations and strong spin-orbit coupling are competing in a 
multi-orbital environment (such as the $5d$ electron iridates compounds). Here the plain
Hubbard model is not sufficient to capture the physics and more
advanced model such as the Kanamori-type Hamiltonian would be the
starting point.\cite{Puetter12} In these situations, more versatile numerical techniques 
are needed -- the hybridization expansion continuous-time QMC cluster dynamical mean
field framework is a promising example among several others.
But just as the presented QMC studies of the KMH model in this review, a
combination of accurate, controlled numerical techniques and clear theoretical
understanding can indeed facilitate controlled investigations of novel
physics in correlated topological systems.

\section*{Acknowledgments}

We thank Fakher Assaad, Victor Chua, Xi Dai, Andrew Essin, Gregory Fiete, 
Zheng-Cheng Gu, Victor Gurarie, Martin Hohenadler, Alejandro Muramatsu, Lei Wang, 
Zhong Wang and Stefan Wessel for collaboration and discussions. HHH 
and ZYM are grateful for the hospitality from Institute for Advanced Study, Tsinghua 
University and Institute of Physics, Chinese Academy of Sciences. ZYM acknowledges the supported by the
NSERC, CIFAR, and Centre for Quantum Materials at the
University of Toronto. HHH acknowledges 
the support by Grant No. ARO W911NF-09-1-0527, Grant No. NSF DMR-0955778, Grant No. 
ARO W911NF-12-1-0573 with funding from the DARPA OLE Program and computer time at 
Texas Advanced Computing Center at the University of Texas, Austin and the Brutus 
cluster at ETH Z\"{u}rich. TCL acknowledges JARA-HPC and JSC J\"{u}lich for the 
allocation of CPU time.


\begin{thebibliography}{99}

\bibitem{Klitzing80}
K. v.~Klitzing, G. Dorda, and M. Pepper, {\it Phys. Rev. Lett.} {\bf 45} (1980) 494.

\bibitem{Halperin82}
B.~I. Halperin, {\it Phys. Rev. B} {\bf 25} (1982) 2185.

\bibitem{Thouless82}
D.~J. Thouless, M. Kohmoto, M.~P. Nightingale, and M. den Nijs, {\it Phys. Rev. Lett.} {\bf 49} (1982) 405.

\bibitem{Avron83}
J. E. Avron, R. Seiler, and B. Simon, {\it Phys. Rev. Lett.} {\bf 51} (1983) 51.

\bibitem{Wen2013}
X.-G. Wen, {\it arXiv}:1301.7675.

\bibitem{Kane05a}
C.~L. Kane and E.~J. Mele, {\it Phys. Rev. Lett.} {\bf 95} (2005) 146802.

\bibitem{Kane05b}
C.~L. Kane and E.~J. Mele, {\it Phys. Rev. Lett.} {\bf 95} (2005) 226801.

\bibitem{Xu2006}
X.~C. Xu and J.~E. Moore, {\it Phys. Rev. B} {\bf 73} (2006) 045322.

\bibitem{Qi2006}
X.-L. Qi, Y.-S. Wu, and S.-C. Zhang, {\it Phys. Rev. B} {\bf 74} (2006) 085308.

\bibitem{Bernevig2006}
B.~A. Bernevig, T.~L. Hughes, and S.-C. Zhang, {\it Science} {\bf 314} (2006) 1757.

\bibitem{Koenig2007}
M. K\"onig, S. Wiedmann, C. Br\"une, A. Roth, H. Buhmann, L.~W. Molenkamp, X.-L. Qi, and S.-C. Zhang, {\it Science} {\bf 318} (2007) 766.

\bibitem{Fu2007prb}
L. Fu and C.~L. Kane, {\it Phys. Rev. B} {\bf 76} (2007) 045302.

\bibitem{Fu2007b}
L. Fu, C.~L. Kane, and E. J. Mele, {\it Phys. Rev. Lett.} {\bf 98} (2007) 106803.

\bibitem{Zhang2009}
H.~J. Zhang, L.~C. Xing, X.-L. Qi, X. Dai, F. Zhong, S.-C. Zhang, {\it Nature Physics} {\bf 5} (2009) 438.

\bibitem{Chen2009}
Y.~L. Chen, J.~G. Analytis, J.-H. Chu, Z.~K. Liu, S.-K. Mo, X.-L. Qi, H.~J. Zhang, D.~H. Lu, X. Dai, Z. Fang, S.~C. Zhang, I.~R. Fisher, Z. Hussain, Z.-X. Shen, {\it Science} {\bf 325} (2009) 178.

\bibitem{Roy2013}
B. Roy and I.~F. Herbut, {\it Phys. Rev. B} {\bf 88} (2013) 045425.

\bibitem{konig2008review}
M. K\"onig, H. Buhmann, L.~W. Molenkamp, T. Hughes, C.-X. Liu, X.-L. Qi, and S.-C. Zhang, {\it J. Phys. Soc. Jpn.} {\bf 77} (2008) 031007.

\bibitem{maciejko2011}
J. Maciejko, T.~L. Hughes, and S.-C. Zhang, {\it Annu. Rev. Condens. Matter Phys.} {\bf 2} (2011) 31.

\bibitem{Wu06}
C. Wu, B. A. Bernevig, and S.-C. Zhang, {\it Phys. Rev. Lett.} {\bf 96} (2006) 106401.

\bibitem{Rachel2010}
S. Rachel and K. Le~Hur, {\it Phys. Rev. B} {\bf 82} (2010) 075106.

\bibitem{vaezi2012}
A. Vaezi, M. Mashkoori, and M. Hosseini, {\it Phys. Rev. B} {\bf 85} (2012) 195126.

\bibitem{yamaji2011}
Y. Yamaji and M. Imada , {\it Phys. Rev. B} {\bf 83} (2011) 205122.

\bibitem{wuwei2012}
W. Wu, S. Rachel, W.-M. Liu, and K. Le Hur, {\it Phys. Rev. B} {\bf 85} (2012) 205102.

\bibitem{shunliyu2011}
S.-L. Yu, X.~C. Xie, and J.-X. Li, {\it Phys. Rev. Lett.} {\bf 107} (2011) 010401.

\bibitem{Lang2011}
M. Hohenadler, T.~C. Lang, and F.~F. Assaad, {\it Phys. Rev. Lett.} {\bf 106} (2011) 100403.

\bibitem{Zheng2011}
D. Zheng, G.~M. Zhang, and C. Wu, {\it Phys. Rev. B} {\bf 84} (2011) 205121.

\bibitem{Hohenadler2012}
M. Hohenadler, Z.~Y. Meng, T.~C. Lang, S. Wessel, A. Muramatsu, and F.~F. Assaad, {\it Phys. Rev. B} {\bf 85} (2012) 115132.

\bibitem{moore08}
J. E. Moore. {\it Nature} {\bf 452} (2008) 970.

\bibitem{Hasan2010}
M.~Z. Hasan and C.~L. Kane, {\it Rev. Mod. Phys.} {\bf 82} (2010) 3045.

\bibitem{Qi2010phystoday}
X.-L. Qi and S.-C. Zhang, {\it Physics Today} {\bf 63} (2010) 33.

\bibitem{Qi2011}
X.-L. Qi and S.-C. Zhang, {\it Rev. Mod. Phys.} {\bf 83} (2011) 1057.

\bibitem{hasan2011}
M.~Z. Hasan and J.~E. Moore, {\it Annu. Rev. Condens. Matter Phys.} {\bf 2} (2011) 55.

\bibitem{Fiete2012}
G.~A. Fiete, V. Chua, M. Kargarian, R. Lundgren, A. R\"uegg, J. Wen, and V. Zyuzin {\it Physica E} {\bf 44} (2012) 845.

\bibitem{Hohenadler2013}
M. Hohenadler and F.~F. Assaad, {\it J. Phys.: Condens. Matter} {\bf 25} (2013) 143201.

\bibitem{xidai2012}
H. Weng, Xi Dai, and Zhong Fang, {\it Asia Pac. Phys. Newslett.} {\bf 1} (2012) 31.

\bibitem{ando2013}
Y. Ando, {\it J. Phys. Soc. Jpn.} {\bf 82} (2013) 102001.

\bibitem{Assaad2013}
F.~F. Assaad, M. Bercx, and M. Hohenadler, {\it Phys. Rev. X} {\bf 3} (2013) 011015.

\bibitem{Fu2006prb}
L. Fu and C.~L. Kane, {\it Phys. Rev. B} {\bf 74} (2006) 195312.

\bibitem{Hung2013b}
H.-H. Hung, V. Chua, L. Wang, and G.~A. Fiete, {\it arXiv}:1307.2659.

\bibitem{Wang2012prx}
Z. Wang and S.-C. Zhang, {\it Phys. Rev. X} {\bf 2} (2012) 031008.

\bibitem{Hung2013}
H.-H. Hung, L. Wang, Z.-C. Gu and G.~A. Fiete, {\it Phys. Rev. B} {\bf 87} (2013) 121113(R).

\bibitem{Lang2013}
T.~C. Lang, A.~M. Essin, V. Gurarie, and S. Wessel, {\it Phys. Rev. B} {\bf 87} (2013) 205101.

\bibitem{Haldane1988}
F.~D.~M. Haldane, {\it Phys. Rev. Lett.} {\bf 61} (1988) 2015.

\bibitem{CastroNeto2009}
A.~H. Castro Neto, F. Guinea, N.~M.~R. Peres, K.~S. Novoselov, and A.~K. Geim, {\it Rev. Mod. Phys.} {\bf 81} (2009) 109.

\bibitem{Kane2007}
C.~L. Kane, {\it Int. J. Mod. Phys. B} {\bf 21} (2007) 1155.

\bibitem{Ezawa2013}
M. Ezawa, Y. Tanaka, and N. Nagaosa, {\it arXiv}:1307.7347.

\bibitem{yang2013}
Y. Yang, H. Li, L. Sheng, R. Shen, D.~N. Sheng, and D.~Y. Xing, {\it arXiv}:1301.1618.

\bibitem{rachel2013}
S. Rachel, {\it arXiv}:1310.3159.

\bibitem{Sheng2005}
L. Sheng, D.~N. Sheng, C.~S. Ting, and F.~D.~M. Haldane, {\it Phys. Rev. Lett.} {\bf 95} (2005) 136602.

\bibitem{Sheng2006}
D.~N. Sheng, Z.~Y. Weng, L. Sheng, and F.~D.~M. Haldane, {\it Phys. Rev. Lett.} {\bf 97} (2006) 036808.

\bibitem{Goth2013}
F. Goth, D.~J. Luitz, F.~F. Assaad, {\it Phys. Rev. B} {\bf 88} (2013) 075110.

\bibitem{Assaad2002}
F.~F. Assaad, {\it NIC Series Vol.} {\bf 10} (2002).

\bibitem{Assaad2008}
F.~F. Assaad and H.~G. Evertz, {\it Lect. Notes Phys.} {\bf 739} (2008) 277.

\bibitem{Sugiyama1986}
G. Sugiyama, S.~E. Koonin, {\it Ann. Phys.} {\bf 168} (1986) 1.

\bibitem{Sorella1989}
S. Sorella, S. Baroni, R. Car, and M. Parrienllo, {\it Europhys. Lett.} {\bf 8} (1989) 663.

\bibitem{White1989}
S.~R. White, D.~J. Scalapino, R.~L. Sugar, E.~Y. Loh, and J.~E. Gubernatis, {\it Phys. Rev. B} {\bf 40} (1989) 506.

\bibitem{Meng2010}
Z.~Y. Meng, T.~C. Lang, S. Wessel, F.~F. Assaad, and A. Muramatsu, {\it Nature} {\bf 464} (2010) 847.

\bibitem{Sorella2012}
S. Sorella, Y. Otsuka, and S. Yunoki, {\it Sci. Rep.} {\bf 2} (2012) 992.

\bibitem{Assaad1996}
F.~F. Assaad and M. Imada, {\it J. Phys. Soc. Jpn.} {\bf 65} (1996) 189.

\bibitem{Feldbacher2001}
M. Feldbacher and F.~F. Assaad, {\it Phys. Rev. B} {\bf 63} (2001) 073105.

\bibitem{Assaad2013b}
F.~F. Assaad and I.~F. Herbut, {\it Phys. Rev. X} {\bf 3} (2013) 031010.

\bibitem{Wang2012a}
Z. Wang, X.-L. Qi, and S.-C. Zhang, {\it Phys. Rev. B} {\bf 85} (2012) 165126.

\bibitem{Ran2008}
Y. Ran, A. Vishwanath, and D.-H. Lee, {\it Phys. Rev. Lett.} {\bf 101} (2008) 086801.

\bibitem{Qi2008}
X.-L. Qi and S.-C. Zhang, {\it Phys. Rev. Lett.} {\bf 101} (2008) 086802.

\bibitem{Wang2010}
Z. Wang, X.-L. Qi, and S.-C. Zhang, {\it Phys. Rev. Lett.} {\bf 105} (2010) 256803.

\bibitem{qithuges2008prb}
X.-L. Qi, T.~L. Hughes, and S.-C. Zhang, {\it Phys. Rev. B} {\bf 78} (2008) 195424.

\bibitem{Wang2012c}
Z. Wang and S.-C. Zhang, {\it Phys. Rev. B} {\bf 86} (2012) 165116.

\bibitem{Wang2013JPC}
Z. Wang and B. Yan, {\it J. Phys.: Condens. Matter} {\bf 25} (2013) 155601.

\bibitem{Gurarie2011}
V. Gurarie, {\it Phys. Rev. B} {\bf 83} (2011) 085426.

\bibitem{niu1985}
Q. Niu, D.~J. Thouless, and Y.-S. Wu, {\it Phys. Rev. B} {\bf 31} (1985) 3372.

\bibitem{Manmanna12}
S.~R. Manmana, A.~M. Essin, R.~M. Noack, and V. Gurarie, {\it Phys. Rev. B} {\bf 86} (2012) 205119.

\bibitem{Budich12a}
J.~C. Budich, R. Thomale, G. Li, M. Laubach, and S.-C. Zhang, {\it Phys. Rev. B} {\bf 86} (2012) 201407.

\bibitem{Werner13}
J. Werner, F.~F. Assaad, {\it Phys. Rev. B} {\bf 88} (2013) 035113.

\bibitem{Go12}
A. Go, W. Witczak-Krempa, G.~S. Jeon, K. Park, and Y.~B. Kim, {\it Phys. Rev. Lett.} {\bf 109} (2012) 066401.

\bibitem{wang2013wavefunction}
Z. Wang and S.~C. Zhang, {\it arXiv}:1308.4900.

\bibitem{Witczak-Krempa13}
W. Witczak-Krempa, G. Chen, Y.~B. Kim, L. Balents, {\it arXiv}:1305.2193.

\bibitem{Puetter12}
C.~M. Puetter, H.~Y. Kee, {\it Europhys. Lett.} {\bf 98} (2012) 27010.

\end{thebibliography}
\end{document}